\documentclass[atoms,article,accept,moreauthors,pdftex,12pt,a4paper]{mdpi}

\lastpage{x}
\doinum{10.3390/------}
\pubvolume{xx}
\pubyear{2015}
\history{Received: xx / Accepted: xx / Published: xx}

\usepackage{graphicx}
\usepackage{epsfig}
\usepackage{psfrag}
\usepackage{amsmath,amssymb}
\usepackage{xcolor}
\usepackage[normalem]{ulem}  
\usepackage{soul}

\newcommand{\copa}{\ensuremath{\hat{a}^\dagger}}

\newcommand{\dopa}{\ensuremath{\hat{a}}}

\newcommand{\cPsi}{\ensuremath{\hat{\Psi}^\dagger}}
\newcommand{\dPsi}{\ensuremath{\hat{\Psi}}}

\newcommand{\av}[1]{\ensuremath{\langle #1 \rangle}}

\newcommand{\<}{\langle}
\renewcommand{\>}{\rangle}
\renewcommand{\(}{\left(}
\renewcommand{\)}{\right)}
\renewcommand{\[}{\left[}
\renewcommand{\]}{\right]}
\newcommand{\beq}{\begin{equation}}
\newcommand{\eeq}{\end{equation}}
\renewcommand{\d}{\dag}
\newcommand{\h}{\hat}

\newcommand{\ej}{\epsilon_j}

\newcommand{\f}{\frac}
\newcommand{\half}{\hbox{$1\over2$}}

\newcommand{\EQREF}[1]{Eq.~(\ref{#1})}
\newcommand{\BEQ}{\begin{equation}}
\newcommand{\EEQ}{\end{equation}}
\newcommand{\BEQA}{\begin{eqnarray}}
\newcommand{\EEQA}{\end{eqnarray}}
\newcommand{\BEQAL}{\begin{align}}
\newcommand{\EEQAL}{\end{align}}

\newcommand{\commentout}[1]{{}}

\Title{Cavity quantum electrodynamics of continuously monitored Bose-condensed atoms}

\Author{Mark D.~Lee and Janne Ruostekoski}

\address[1]{%
Mathematical Sciences, University of Southampton, Southampton SO17
1BJ, United Kingdom}

\corres{janne@soton.ac.uk}

\abstract{
We study cavity quantum electrodynamics of Bose-condensed atoms that are subjected
to continuous monitoring of the light leaking out of the cavity. Due to a given
detection record of each stochastic realization, individual runs spontaneously
break the symmetry of the spatial profile of the atom cloud and this symmetry can
be restored by considering ensemble averages over many realizations. We show that
the cavity optomechanical excitations of the condensate can be engineered to
target specific collective modes.  This is achieved by exploiting the spatial
structure and symmetries of the collective modes and light fields. The cavity
fields can be utilized both for strong driving of the collective modes and for
their measurement. In the weak excitation limit the condensate-cavity system may
be employed as a sensitive phonon detector which operates by counting photons
outside the cavity that have been selectively scattered by desired phonons.
}



\begin{document}



\section{Introduction}

Cavity quantum electrodynamics (cQED) is a paradigm model of quantum
optics~~\cite{CarmichaelVol2}. In typical realizations a single atom interacts
strongly with a single quantized light field. A large coupling
coefficient $g$ of an atom to the cavity field, as compared with the cavity and
spontaneous decay rates, then ensures a cooperative coupling regime. Recently
it has also become possible to confine quantum degenerate atoms inside a single mode
high-finesse optical
cavity~\cite{Brennecke2007a,Colombe2007a,Murch2008a,Brahms2012a,Botter2013a,Zimmermann,Hemmerich},
representing an experimental milestone in extending quantum optical systems to
the realm of quantum many-atom physics~\cite{Ritsch2013a}. For example, atomic
Bose-Einstein condensates (BECs) can be strongly coupled to an optical cavity
field already at a single photon level.

The idea of optomechanics, the use of light forces to control and manipulate the
quantum properties of mechanical oscillators, can be realized for the case of
BECs in a cavity~\cite{Brennecke2008a,Murch2008a,Brahms2012a,Botter2013a}. Here,
typically, the motion of the condensate is employed as a mechanical device that
is coupled to the cavity light field. In the case of a BEC, the oscillator is
already in the ground state and there is no need for additional cooling.
Optomechanics addresses  the fundamental interest of oscillators operating in
the  quantum regime and is promising for highly sensitive measurements of weak
forces at the quantum limit.

One of the central elements of quantum optics and quantum physics is the effect
of quantum measurement on the evolution of a quantum system. A continuous
measurement process forms a coupling of the quantum system to an environment.
The resulting open quantum system dynamics is conditioned on the particular
measurement record in each experimental run. In open interacting quantum
many-body systems the studies of continuous monitoring of the system evolution
and the backaction of the quantum measurement process pose considerable
challenges. For small quantum systems, the full quantum treatment of the
measurement backaction may be incorporated in stochastic master equations
and stochastic quantum trajectories of state vectors (quantum Monte Carlo wave
functions)~\cite{Tian1992a,Dalibard1992a,Dum1992a}. These approaches can produce
a faithful representation of a possible measurement record for an individual
experimental run, where the dynamics is conditioned on the stochastic
measurement outcomes. For very large systems quantum trajectory simulations are
not possible and there is a quest to develop approximate computationally
efficient approaches. The motive for such developments is, e.g., the observation
of the measurement backaction of the cavity output on the dynamics of the atoms
in ultracold atom experiments~\cite{Murch2008a,Brahms2012a}. Outside the
ultracold regime, a continuous measurement process of the cavity output has been
employed in the preparation of spin squeezed atomic
ensembles~\cite{spinsqueezing_vuletic}.

Here we investigate the cavity optomechanics of a BEC, especially focusing on
single stochastic realizations of the dynamics that are continuously monitored
by detecting the light leaking out of the cavity. In the numerical simulations
the BEC-cavity dynamics is integrated using classical stochastic measurement
trajectories~\cite{measuretraj} that are based on stochastic differential
equations (SDEs) for an approximate phase-space representation of a continuously
monitored multimode system~\cite{measuretraj,Javanainen2013a}. The advantage of
the approximate treatment is that nonlinear evolution on a spatial grid of over
a thousand points can be implemented and there is, e.g., no need for linearizing
about a mean-field steady-state
solution~\cite{Gardiner2001a,Nagy2008a,Szirmai2009a,odell2}.

Owing to the multimode nature of the BEC, the light not only excites the
center-of-mass motion but also other collective modes of the condensate. The
detection of light outside the cavity represents a combined excitation and
measurement of the collective BEC modes. By exploiting the spatial structure and
symmetries of the collective modes and light fields we can engineer the light
excitation specifically to target certain modes. The driving of modes may also
be achieved at higher light intensities but this is shown to considerably
enhance the phase decoherence rate. The systems that are solely driven by the
measurement of light outside the cavity are of special interest and represent a
quantum measurement-induced spontaneous symmetry breaking: each stochastic
realization exhibits a characteristic evolution of the density pattern, but
ensemble-averaging over many runs restores the stationary unbroken spatial
profile.

We show that, in the weak excitation limit, the selective coupling of the
collective modes by the tailored BEC-cavity system may be employed as a
sensitive phonon detector. The phonon detector is based on a photon counting of
light leaking outside the cavity in which case the light that is not scattered
by the phonons is suppressed by interference. The technique could potentially
open the gate for a sensitive single phonon detector of BEC excitations that
could measure statistical properties of the phonons, act as an accurate BEC
thermometer, and prepare complex quantum states of phonons.

\section{Basic model}
\subsection{Hamiltonian formalism and open system dynamics}

\begin{figure}
\center{
\epsfig{file=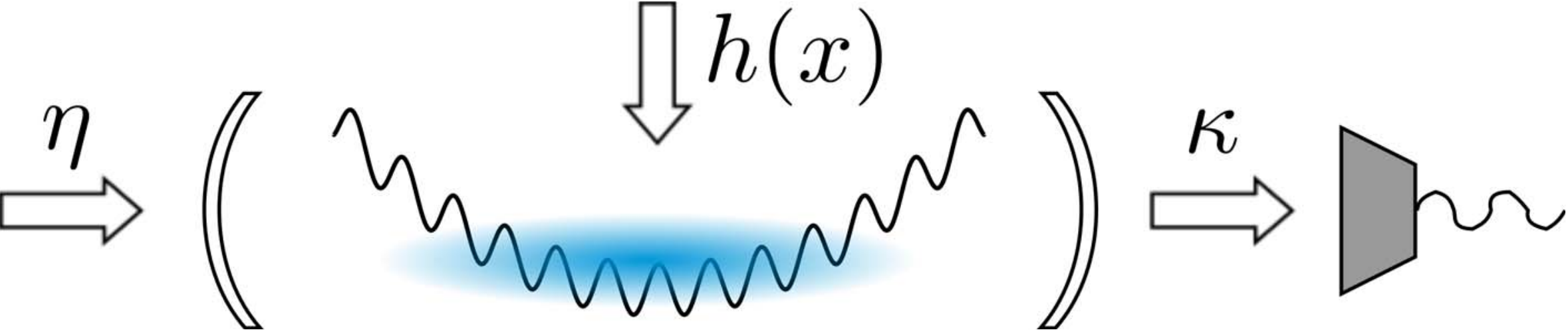,width=0.8\columnwidth}
\caption{Schematic of the cavity-BEC system.  The single-mode cavity can be pumped
on axis at rate $\eta$, while the BEC inside the cavity can be pumped directly by
a transverse beam of profile $h(x)$. The photons leaking from the cavity (at a rate
$2\kappa$) are continuously monitored. The atoms are subject to an external
trapping potential $V(x)$ that is illustrated by the black line. Here the potential is shown
mimicking a superposition of a periodic
optical lattice and a harmonic trap, but generally the potential can take an essentially arbitrary form.}
\label{fig:StrongLatticeSystem}}
\end{figure}

In the following we introduce the formalism for the BEC-cavity system, illustrated
in Fig.~\ref{fig:StrongLatticeSystem}, where the photons are leaking out of the
cavity through the cavity mirrors. In the rotating wave approximation (and in the
rotating frame of the pump field) we write the Hamiltonian for the closed system
of two-level atoms and the cavity in the second-quantized
form~\cite{Jaynes1963a,WallsMilburn,Maschler2008a}
\BEQ
H_{ge} = H_{A}+H_C+H_{CA},
\EEQ
where the Hamiltonian terms for the atoms, $H_A$, for the cavity, $H_C$, and for the
atom-cavity coupling, $H_{CA}$, are written as
\begin{align}
H_A &= \int d x
\cPsi_g(x)H_0^{(g)}\dPsi_g(x) +\frac{U}{2}\int d x \cPsi_g(x)\cPsi_g(x)\dPsi_g(x)\dPsi_g(x) \nonumber \\
&+\int dx
\cPsi_e(x)\left[H_0^{(e)}-\hbar\Delta_{pa}\right]\dPsi_e(x)
-i\hbar\int dx
\left[\cPsi_g(x)h(x)\dPsi_e(x)-\cPsi_e(x)h(x)\dPsi_g(x)\right] , \\
H_0^{(j)} &= -\frac{\hbar^2 \nabla^2}{2m}+V^{(j)}(x),\\
H_{CA} &=  -i\hbar\int dx \cPsi_g(x) g(x) \copa \dPsi_e(x) + \mbox{H.c.}
\, ,\\
H_A &= -\hbar\Delta_{pc}\copa\dopa+i\hbar\eta(\copa-\dopa)\, .
\end{align}
Here $\omega_c$ and $\omega_a$ denote the cavity and atom resonance frequencies, respectively.
An essentially arbitrary external trapping potential is denoted by $V(x)$.
For the atomic fields $\dPsi_{g(e)}(x)$ annihilates an atom in the ground (excited) state at
position $x$, while $\dopa$ annihilates a photon from the single cavity mode.
The atom-cavity coupling is described by
\BEQ
g(x) = g_0\sin(kx)\,.
\EEQ
The cavity field is pumped along the cavity axis at a rate $\eta$ and via a
transverse beam of profile $h(x)$. In practise, we consider situations where only one
of the two driving mechanisms is used.

We assume that the atoms are tightly confined in a 1D cigar-shaped trap that is oriented
along the cavity axis and neglect any
density fluctuations of the atoms along the radial directions of the trap.
The interatomic interactions for the ground-state atoms are therefore represented by
a 1D interaction strength $U = 2\hbar\omega_\perp a_s$, where $a_s$ denotes
the $s$-wave scattering length and $\omega_\perp$ the radial trapping frequency for the atoms (the confinement perpendicular to the cavity field).
The trapping potential along the axial direction is $V^{(j)}(x)$, where $j$ refers to either
the excited- or the ground-state atoms.
The detunings between the pump frequency $\omega_p$ and the cavity and the atomic resonance
frequencies are denoted by $\Delta_{pc} =
\omega_p-\omega_c$ and
$\Delta_{pa} = \omega_p-\omega_a$, respectively.

We will simplify the system description by considering a large detuning limit $\Delta_{ca}\gg \kappa$ (we also assume that $\Delta_{pa}$ is large so that the spontaneous emission to the modes other than the cavity mode may be ignored), where the excited-state
atomic field may be adiabatically eliminated. The dynamics may then be obtained from the effective Hamiltonian
\begin{align}
H_1 &= \int \mbox{d}x \cPsi(x)\Bigg\{ H_0+\frac{\hbar}{\Delta_{pa}}\bigg[h(x)^2+g(x)^2\copa\dopa
+h(x)g(x)\left(\dopa+\copa\right)\bigg]\Bigg\}\dPsi(x) \nonumber \\
&+ \frac{U}{2}\int \mbox{d}x \cPsi(x)\cPsi(x)\dPsi(x)\dPsi(x) -\hbar\Delta_{pc}\copa\dopa-i\hbar\eta\left(\dopa-\copa\right)\,. \label{eq:Heff}
\end{align}
Here $H_0$ now only refers to the ground-state atoms and we drop the corresponding subscript.

So far, we have only described the closed system of the driven BEC and the cavity.
The open-system description follows from the fact that cavity photons are leaking through
the cavity mirrors at a rate $2\kappa$. In order to analyze a continuous measurement process
of photons outside the cavity, we will assume that all the photons
leaked out of the cavity are detected.
The density operator $\rho_\mathrm{tot}$ for the  BEC-cavity system then evolves according to the master equation
\BEQ
\frac{d \rho_\mathrm{tot}(t)}{d t} = -\frac{i}{\hbar}\left[
H_1,\rho_\mathrm{tot}(t)\right]+\kappa {\cal L}[\dopa]\rho_\mathrm{tot}(t),
\label{eq:mastereqn}
\EEQ
where the superoperator ${\cal L}[\hat O]$ acting on $\hat O$  is defined by
\BEQ
{\cal L}[\hat O]\rho \equiv \left(2\hat O\rho\hat O^\dagger-\hat O^\dagger\hat O\rho-\rho\hat O^\dagger \hat O\right).
\EEQ

The next level of simplification for the dynamics can be obtained by
also adiabatically eliminating the cavity light mode. This can be done in the bad
cavity limit $\kappa \gg N g_0^2/\Delta_{pa}$. The elimination is performed for
the open system dynamics by means of unitary transformations, but we give here
only a simple heuristic explanation of the derivation (for a more rigorous
elimination, a reader is referred to Ref.~\cite{measuretraj}). The equation of
motion for $\dopa$ reads
\BEQ
\frac{d{\dopa}}{dt} =
\left(i\left[\Delta_{pc}-\frac{1}{\Delta_{pa}}\int\cPsi(x)\dPsi(x)g^2(x)dx\right]-\kappa\right)\dopa
+\eta-\frac{i}{\Delta_{pa}}\int\cPsi(x)\dPsi(x)h(x)g(x)dx\,.
\EEQ
The cavity field is then eliminated by setting
\BEQ
\dopa =
\frac{1}{\kappa-i\tilde{\Delta}_{pc}}\left(\eta-\frac{i}{\Delta_{pa}}\int\cPsi(x)\dPsi(x)h(x)g(x) dx\right),\quad \tilde{\Delta}_{pc} \equiv
\Delta_{pc}-\frac{1}{\Delta_{pa}}\int\cPsi(x)\dPsi(x)g^2(x)dx,
\EEQ
We expand the denominator in terms of the small parameter $\tilde{\Delta}_{pc}/\kappa$ (valid when $\kappa\gg \Delta_{pc}, N g_0^2/\Delta_{pa}$),
leading to
\BEQ
\dopa \simeq
\frac{1}{\kappa}\left(\eta-\frac{i}{\Delta_{pa}}\int\cPsi(x)\dPsi(x)h(x)g(x)dx\right) \left[1+i\frac{\tilde{\Delta}_{pc}}{\kappa}+O\left(\frac{\tilde{\Delta}_{pc}^2}{\kappa^2}\right)\right]\,.
\label{eq:AdElima}
\EEQ
For the case that we only pump transversely (with a  beam profile
$h(x)$) we set $\eta=0$. Then
\BEQ
\dopa \simeq
-i\frac{\hat{Y}}{\kappa}\left(1+i\frac{\Delta_{pc}}{\kappa}-i\frac{\hat{X}}{\kappa}\right)\,,
\label{eq:tradel}
\EEQ
where $\hat{Y}$ represents the off-resonant excitation of the atoms
via the transverse pump beam and $\hat{X}$ excitation via the cavity field
\BEQ
\hat{Y} \equiv \int dx \frac{h(x)g(x)}{\Delta_{pa}}\cPsi(x)\dPsi(x),\quad \hat{X} \equiv \int dx \frac{g^2(x)}{\Delta_{pa}}\cPsi(x)\dPsi(x).
\label{eq:Ydefn}
\EEQ
Eliminating the light field in the lowest order approximation [only keeping the
first term in \EQREF{eq:tradel}]  then results in an effective Hamiltonian and the
master equation
\begin{align}
H_{\rm tra} &= \int \mbox{d}x \cPsi(x)\Bigg[ H_0 +\frac{U}{2}\cPsi(x)\dPsi(x)
+\hbar\frac{h^2(x)}{\Delta_{pa}}\Bigg]\dPsi(x),\\
\frac{d \rho_a(t)}{d t} &= -\frac{i}{\hbar}\left[
H_{\rm tra},\rho_a(t)\right]+\frac{1}{\kappa}{\cal L}[\hat{Y}]\rho\,,
\label{eq:mastereqnatomtransverse}
\end{align}
As the light field is eliminated, the measurement observable now depends solely
on atomic operators~\cite{measuretraj}.
In particular, the measurement operator involves
an integral over a nonuniform multimode quantum field $\dPsi(x)$ combined with
a spatially varying pump profile and the cavity coupling strength.
The rate of measurement is given by
\BEQ
r_\mathrm{meas}(t) =
\frac{2}{\kappa}\av{\hat{Y}\hat{Y}}\label{eq:measratetrans} = 2\kappa n,
\EEQ
which we have expressed in terms of the number of photons in the
cavity $n = \av{\copa\dopa} = \av{(\hat{Y}/\kappa)^2}$.  Here all cavity photons appear from interactions of the transverse beam with
atoms and the rate of measurement events which affect the atoms $r_\mathrm{meas}$
is therefore simply that of the number of photons leaving the cavity.

For the case that the cavity mode is driven axially and there is no transverse
pumping of the atoms ($h(x)=0$), eliminating the cavity field operator leads, in
the lowest order in our expansion parameter, to the Hamiltonian and master
equation for the atoms
\begin{align}
H_{\rm axi} &= \int \mbox{d}x \cPsi(x)\Bigg[H_0+\frac{U}{2}\cPsi(x)\dPsi(x)
+\hbar\frac{|\eta|^2}{\kappa^2}\frac{g^2(x)}{\Delta_{pa}}\Bigg]\dPsi(x),
\label{eq:Heff_atom_cavitypump}\\
\frac{d \rho_a(t)}{d t} &= -\frac{i}{\hbar}\left[
H_{\rm axi},\rho_a(t)\right]+\frac{|\eta|^2}{\kappa^3}{\cal L}[\hat{X}]\rho_a(t).
\label{eq:mastereqnatom}
\end{align}
and the rate
of scattered photons counted by the measurement apparatus is
\BEQ
r_\mathrm{meas}(t) =
\frac{2|\eta|^2}{\kappa^3}\av{\hat{X}\hat{X}}.\label{eq:measrate}
\EEQ

The master equations \eqref{eq:mastereqn}, \eqref{eq:mastereqnatomtransverse}, and \eqref{eq:mastereqnatom} represent an ensemble average over a large number of measurement realizations and are not conditioned on any particular measurement record. They do not incorporate information
about individual stochastic runs. A common approach to describe the backaction of a continuous quantum measurement is to unravel the master equation into quantum trajectories of the state vectors~\cite{Tian1992a,Dalibard1992a,Dum1992a,Carmichael1993a}. A full quantum treatment of quantum trajectory simulations is computationally demanding in large systems and in the following we will develop approximate methods for individual stochastic runs.

\subsection{Phase-space and stochastic descriptions}
\label{sec:classicalphasespace}

Typical BEC-cavity systems may consist of spatial atom dynamics that require well
over 100 modes for an accurate description. On the other hand, the atom number in
the cavity may commonly vary between $10^3$-$10^6$. This is a significantly larger
system than the full quantum description would allow in a numerical simulation.
Here we introduce an approximate computationally feasible approach that is based
on classical phase-space methods. The state of the multimode atom-light system can
be expressed by the Wigner function $W(\alpha,\alpha^*,\{\psi,\psi^*\})$, where
$\alpha$ is the classical variable associated with $\dopa$ and $\psi$ is a
classical field representation of the field operator $\dPsi$ that is
stochastically sampled from an ensemble of Wigner distributed classical fields.
The full quantum dynamics of the master equation can be mapped to a phase-space
dynamics of the Wigner function using standard techniques of quantum
optics~\cite{WallsMilburn,QuantumNoise}. Neither the master equation for the
density matrix nor the phase-space dynamics for the Wigner function are conditioned
on any particular measurement record, but represent an ensemble average over a
large number of measurement realizations. In order to describe a single
experimental run of a continuously monitored BEC-cavity system, where the photons
leaking out of the cavity are measured  and the state of the system is determined
by the detection record, we need to have an alternative representation to the
ensemble-averaged ones. We will therefore derive classical stochastic measurement
trajectories from an approximate description of the phase-space dynamics. Each
such a trajectory is a faithful representation of a single experimental run where
the dynamical noise of the equations is associated with the measurement noise and
the detection record of the output light from the cavity.

The equation of motion for the Wigner function of the BEC-cavity system may be
derived from the master equation~(\ref{eq:mastereqn}) via the operator
correspondences~\cite{QuantumNoise,WallsMilburn} similar to
\BEQ
\dPsi\rho \leftrightarrow \left( \psi+\frac{1}{2}\frac{\delta}{\delta\psi^*}\right)W(\alpha,\alpha^*,\{\psi,\psi^*\})\,.
\EEQ
The substitution of the operator correspondences to the master equation leads to a
\textit{Fokker-Planck equation} (FPE) for the Wigner function in the limit of weak
quantum fluctuations~\cite{measuretraj}. Specifically, for the interatom
interactions we consider the limit the atom number $N\rightarrow \infty$, while
keeping $C = N U$ constant. Analogously, for the atom-photon interaction terms we
take the limit where the number of cavity photons $n\rightarrow \infty$  while the
maximum atom-photon interaction energies  $\chi = \hbar(g_0^2/\Delta_{pa})n$ and
$\chi_h =\hbar (h_0g_0/\Delta_{pa})\sqrt{n}$  remain constant (note that the
transverse pump scales as $h_0\propto \sqrt{n}$). The interatomic interaction
limit can be related to the 1D Tonks parameter~\cite{olshanii_98} $\gamma =
mU/(\hbar^2\rho_{1D})\gg 1$, where $\rho_{1D}$ is the one-dimensional atom
density, indicating that the expansion is strictly valid in the regime
of a weakly interacting bosonic gas (when the Bogoliubov approximation becomes
accurate for the ground-state atoms), although especially in 1D systems short-time
behavior can be qualitatively described even for more strongly fluctuating
cases~\cite{RUO05}.
The classical approach can also be significantly more accurate in estimating the dynamics
of the \emph{measured observable} even deep in the quantum regime~\cite{Javanainen2013a}.
The Tonks parameter measures the ratio of the nonlinear
$s$-wave interaction to kinetic energies for atoms spaced at the mean interatomic
distance and the expansion also implies that the number of atoms found within a
healing length $\xi$ is $N_\xi\simeq 1/\sqrt{2\gamma}\gg 1$.

In the limit of weak quantum fluctuations the FPE for the approximate BEC-cavity
system then reads~\cite{measuretraj}
\BEQ
\frac{\partial }{\partial
t}W = -\sum_i\frac{\partial }{\partial q_i} A_i W +\frac{\kappa}{2}\frac{\partial^2}{\partial\alpha\partial\alpha^*}
W+\frac{\kappa}{2}\frac{\partial^2}{\partial\alpha^*\partial\alpha}
W\, ,
\label{eq:FPeqn}
\EEQ
where the index $q_i$ runs over the set
$\big\{\alpha,\alpha^*,\psi(x), \psi^*(x)\big\}$.  The nonlinear atom-light
dynamics is incorporated in the drift term elements $A_i$ that arise from the
unitary Hamiltonian~(\ref{eq:Heff})
\begin{align}
A&_\alpha = -\eta +i \int
\frac{h(x)g(x)}{\Delta_{pa}}|\psi(x)|^2 dx
+\left(\kappa-i\Delta_{pc}+ i\int
\frac{g^2(x)}{\Delta_{pa}}|\psi(x)|^2dx\right)\alpha \\
A&_{\psi(x)}= \frac{i}{\hbar}\int dx \Bigg(H_0+ U|\psi(x)|^2
+\hbar\left[\frac{h^2(x)}{\Delta_{pa}}+\frac{g^2(x)|\alpha|^2}{\Delta_{pa}}+
\frac{h(x)g(x)}{\Delta_{pa}}(\alpha+\alpha^*)\right]\Bigg)\psi(x)
\end{align}
The last two terms in \EQREF{eq:FPeqn} can be physically associated with the
backaction of the continuously measured light of the leaking out of the cavity and
they form the diffusion part of the equation.

In deriving \EQREF{eq:FPeqn} we have neglected the terms containing higher derivatives than the second order ones by taking the weak fluctuation limit. The advantage of expressing the dynamics as a FPE follows from their mathematical correspondence to systems of SDEs~\cite{QuantumNoise,CarmichaelVol1}.
Besides the computational simplicity of SDEs as compared with \EQREF{eq:FPeqn}, we can now also obtain a stochastic description for single realizations of a continuous measurement process. On the other hand, the corresponding  FPE corresponds to an ensemble average over all possible measurement outcomes that has discarded the individual measurement records.

The derivation of a FPE is reminiscent of dropping the triple derivative terms that arise from the $s$-wave interactions in the truncated Wigner approximation~\cite{drummond93}
that has been actively utilized in the studies of bosonic atom dynamics in closed
systems~\cite{Steel1998a,Sinatra2002a,Isella2006a,Blakie2008a,Martin2010a,Polkovnikov2010a,Opanchuk2012a} (for a recent work on using truncated Wigner approximation in cold atoms see, e.g., \cite{Mathey,gross_esteve_11,Plimak,Martin_bright,Opanchuk_inter,Dujardin,Lewis-Swan})
as well as when incorporating three-body losses of atoms in a trap~\cite{Norrie2006b,Opanchuk2012a}.

A FPE can be mathematically mapped to Ito SDEs. For the cavity system we can express \EQREF{eq:FPeqn} as a coupled SDEs for the stochastic light and atomic amplitudes $\alpha(t)$ and  $\psi(x,t)$, respectively. These equations can be numerically integrated even for large systems and they represent an unraveling of the FPE into classical stochastic measurement trajectories~\cite{measuretraj}. The trajectories describe individual continuous measurement processes where the dynamics is conditioned on the detection record.
Here we, however, focus on the specific limit of a bad cavity $\kappa \gg N g_0^2/\Delta_{pa}$, when the cavity light field can be adiabatically eliminated from the dynamics. We consider the transversely pumped case of \EQREF{eq:mastereqnatomtransverse}. Having eliminated the cavity field we can now use a Wigner representation in terms solely of the atomic variables $W(\{\psi(x),\psi^*(x)\})$. The approximate FPE can be derived in the case of weak quantum fluctuations using the same principles as \EQREF{eq:FPeqn}. We find
\begin{equation}
\frac{\partial }{\partial
t}W = \left.\frac{\partial }{\partial
t}W\right|_{\rm Ham} + \left.\frac{\partial }{\partial
t}W\right|_{\rm meas}
\label{simplefpe}
\end{equation}
where the first term corresponds to that from the Hamiltonian evolution governed
by \EQREF{eq:mastereqnatomtransverse}
\begin{equation}
\left.\frac{\partial }{\partial
t}W\right|_{\rm Ham} = \frac{i}{\hbar}\int dx \frac{\partial}{\partial \psi}
\left[H_0 +U|\psi(x)|^2
+\hbar\frac{h^2(x)}{\Delta_{pa}}\right]\psi(x) W +\mbox{c.c.},
\end{equation}
while the measurement term has the form
\begin{align}
\frac{\partial }{\partial t}&W\Bigg|_{\rm meas} =
\frac{1}{\kappa} \int dx
\frac{h^2(x)g^2(x)}{\Delta_{pa}^2} \frac{\delta}{\delta\psi(x)} \psi(x)W
\nonumber \\
&+\frac{1}{\kappa} \int dx
dx'\frac{h(x)g(x)}{\Delta_{pa}(x)}\frac{h(x')g(x')}{\Delta_{pa}(x')}\left(
\frac{\delta^2}{\delta\psi(x)\delta\psi^*(x')} \psi(x)\psi^*(x')-
\frac{\delta^2}{\delta\psi(x)\delta\psi(x')} \psi(x)\psi(x')\right) W\nonumber \\
&+ \mbox{ c.c.}
\end{align}

The diffusion matrix of the FPE for the part that does not involve measurements
vanishes identically.
Symmetrically-ordered expectation values $\<\cdots\>_W$ for the atomic fields are obtained with respect to the
quasidistribution function $W(\{\psi(x),\psi^*(x)\})$
\BEQ
\<\psi^*(x_1)\cdots\psi^*(x_k)\psi(x_{k+1})\cdots\psi(x_l) \>_W=\int d^2\psi\, W(\{\psi,\psi^*\})
\psi^*(x_1)\cdots\psi^*(x_k)\psi(x_{k+1})\cdots\psi(x_l)\,.\label{eq:wignerexp}
\EEQ
The FPE can then be unraveled into classical trajectories for the stochastic field $\psi(x)$ obeying
\BEQ
d\psi(x) =
\Bigg\{\frac{-i}{\hbar}\left[H_0+U|\psi(x)|^2\right]
-i\frac{h^2(x)}{\Delta_{pa}(x)}
-\frac{1}{\kappa}\frac{h^2(x)g^2(x)}{\Delta_{pa}^2(x)}
\Bigg\}\psi(x)dt-i\sqrt{\frac{2}{\kappa}}
\frac{h(x)g(x)}{\Delta_{pa}(x)}\psi(x)dW\,,\label{eq:ElimTruncWignerSDETrans}
\EEQ
where $dW$ denotes a Wiener increment with $\av{dW}=0$,
$\av{dW^2}=dt$.

At first glance, the last term proportional to $dt$ would appear to give rise to
non-unitary evolution.  However, this term counteracts the non-unitary evolution
introduced by the term proportional to $dW$, and total atom number is in fact
conserved by \EQREF{eq:ElimTruncWignerSDETrans}.  These two terms, which are
$\propto h(x) g(x)$, represent the effect of the light detection record on the
atoms.  For those terms which explicitly lead to unitary evolution, that which is
proportional to $h^2(x)$ describes the light shift due to the transverse beam,
while the remainder describe the evolution of the BEC that would occur in the
absence of the cavity mode.

The stochastic noise in \EQREF{eq:ElimTruncWignerSDETrans} is a physical
consequence of the backaction of a continuous quantum measurement process of the
light that has leaked out of the cavity. In this classical approximation to a
single experimental run the dynamics is conditioned on the detection record. The
noise term in the SDE  \eqref{eq:ElimTruncWignerSDETrans} directly results from
the diffusion term in the corresponding FPE. The ensemble average of the dynamics
over many stochastic realizations generates the unconditioned expectation values
\eqref{eq:wignerexp}.
Different multimode treatments of continuously measured systems,
that are based on alternative phase-space approaches
and are also suitable for cavity systems, were developed in Refs.~\cite{Szigeti2009a,Hush2013a}.
The method also has similarities to numerical approaches to `stochastic electrodynamics', see for instance~\cite{Carmichael_SED}.

In the stochastic representation the initial conditions
$W(\{\psi,\psi^*\},t=0)$   correspond to a (Wigner-distributed)
classical probability distribution for the initial state.
Thermal and quantum fluctuations may be included in the initial state of $\psi$ within the
constraint that the corresponding $W(\{\psi,\psi^*\},t=0)$ remains
positive~\cite{twabook}. This still allows notable quantum fluctuations, such as mode and spin squeezing, to be incorporated. In practical situations, for an accurate modeling of short-time dynamics it is
often necessary to sample the initial conditions $\psi(x,t=0)$ of individual
stochastic realizations from $W(\{\psi,\psi^*\},t=0)$ using
many-body theories that sufficiently well reproduce  the correct quantum
statistical correlations for an initially stable equilibrium configuration of the
system. For simplicity, we consider the initial configuration of the atoms in the
ground state inside the cavity in the absence of the light field. The general idea
is to represent the many-body initial state in terms of some non-interacting
quasi-particles $\h\beta_j$, $\h\beta^\d_j$ that satisfy the ideal gas phonon
statistics

\beq
\<\h\beta_j^\dagger \h\beta_j\>=\bar{n}_j={1\over\exp{(
\ej/k_BT)}-1}\,.
\eeq
We then replace the quantum operators $(\h\beta_j,\h\beta_j^\d)$  by the complex stochastic
variables $(\beta_j,\beta_j^*)$, obtained by sampling the
corresponding Wigner distribution of the quasiparticles. The operators $(\h\beta_j,\h\beta_j^\d)$
behave as a collection of ideal harmonic oscillators whose Wigner
distribution in a thermal bath reads~\cite{QuantumNoise}
\beq
\label{wigner} W(\beta_j,\beta_j^*)=\f{2}{\pi}\tanh \( \xi_j\)
\exp\[ -2|\beta_j|^2\tanh\( \xi_j\)\]\,,
\eeq
where $\xi_j\equiv \ej /2k_B T$.  The nonvanishing
contribution to the width $\bar{n}_j+\half$ of the Gaussian distribution at $T=0$ for each mode represents the
quantum noise. Since the Wigner function returns symmetrically
ordered expectation values, we have
\beq
\<\beta_j^*\beta_j\>_W =\int d^2\beta_j\, W(\beta_j,\beta_j^*)
|\beta_j|^2= \bar{n}_j+{1\over2} \,,
\eeq
and similarly $\<\beta_j\>_W=\<\beta_j^*\>_W=\<\beta_j^2\>_W=0$.

In the Bogoliubov approximation we introduce the quasi-particles by expanding the field operator $\h\psi(x,t=0)$ in terms of the BEC ground state amplitude $\h\beta_0\psi_0$, with $\<\h\beta_0^\dagger\h\beta_0\>=N_0$ (here $N_0$ denotes the BEC atom number that excludes the depleted atoms in the excited states), and the excited states~\cite{Isella2006a}
\beq
\label{field} \h\psi(x)=\psi_0(x)\h\beta_0+ \sum_{j>0} \big[
u_j(x)\h\beta_j-v^*_j(x)\h\beta^\d_j \big]\,.
\eeq
The ground-state BEC solution $\psi_0$ ($\int dx |\psi_0(x)|^2 =1$), the quasiparticle mode functions $u_j(x)$ and $v_j(x)$ ($j>0$), and the corresponding eigenenergies $\ej$  can be solved numerically.  In a more strongly
fluctuating case the quasi-particle modes and the ground-state condensate profile
may be solved self-consistently using the Hartree-Fock-Bogoliubov
theory~\cite{gross_esteve_11,Cattani2013a}. A strongly confined 1D system may also
exhibit enhanced phase fluctuations that can be incorporated using a
quasi-condensate representation~\cite{Martin2010a}. Finally, the initial state $\psi(x,t=0)$ for the stochastic simulations that synthesizes the appropriate statistics  may then be constructed from $\h\psi(x,t=0)$ in \EQREF{field} by replacing $(\h\beta_j,\h\beta_j^\d)$ by the stochastically sampled $(\beta_j,\beta_j^*)$.

\section{Numerical results}
\label{sec:optomechanics}

\subsection{Cavity optomechanical system}

The coupling of a BEC to the cavity light mode and the continuous measurement of
light leaking out of the cavity generate a mechanical response on the atoms.
In the field of cavity optomechanics the coupling of a variety of mechanical oscillators to the cavity
field has been actively investigated in recent years~\cite{Kippenberg2007a,Kippenberg2008a,Meystre2013a,Aspelmeyer2013a}.
Nanomechanical resonators have been cooled to the quantum
regime~\cite{Oconnell2010a,Teufel2011a,Chan2011a}, with potential applications, e.g., to
quantum sensing. A BEC in a cavity already forms the ground state where the atomic motion can be coupled to the cavity mode~\cite{Brennecke2008a,Murch2008a,Brahms2010a,Brahms2012a,Botter2013a}.

The coupling of the cavity field to intrinsic collective excitations of a BEC was studied using classical measurement trajectories in Ref.~\cite{measuretraj}. Here we expand those studies and focus in particular on single trajectories that are conditioned on the measurement record on the light outside the cavity.
We show how the system can be engineered in such a way that the light excitation
specifically targets certain modes and how varying the pump power affects the resulting dynamics.
Here we also demonstrate how in the weak excitation limit the
condensate-cavity system may be employed as a sensitive phonon detector which operates by
counting photons outside the cavity that have been selectively scattered by desired phonons.

The continuous  measurements generate optomechanical dynamics of a BEC. Owing to the multimode nature of the BEC, the number of internal dynamical degrees of freedom
of the atoms is large, involving many collective excitation modes. It is especially interesting to consider situations where the collective excitations of the BEC are solely the consequence of the
conditioned measurement record of a single stochastic realization. In such cases the individual trajectories may display a notable variety of different dynamics, but ensemble-averaging
over a large number of realizations then cancels out any overall motion in the
atomic density.

We consider regimes where the excitations of the BEC are not small and they also
exhibit a complex spatial structure, such that standard treatments of the linear
optomechanical regime are insufficient. The phonon excitations are a characteristic signature of
the BEC nonlinearity.
In general, detection of the light
leaking out of the cavity represents a combined measurement and excitation of
several of the interacting collective modes of the BEC. However, we demonstrate
examples of how the measurement can be tailored to preferentially excite and detect
selected intrinsic excitations.

For simplicity, we consider a BEC confined in a
harmonic well inside an optical cavity and set
$
V(x) = m\omega^2x^2/2
$, with corresponding length scale $x_0 = \sqrt{\hbar/(m\omega)}$ and time scale
$t_0 = 1/\omega$.
The wavelength of the cavity is assumed to be of the same order as the size of the
BEC.  This system also represents the translationally invariant case of many
BECs in a periodic potential with negligible tunneling, with each BEC coupled
identically.

In the simulation examples we consider the limit of weak
quantum fluctuations and the limit where the light field can be adiabatically
eliminated ($n\gg1$ and $ N g_0^2/\Delta_{pa}\ll \kappa$), corresponding to
\EQREF{eq:ElimTruncWignerSDETrans}. The SDE \eqref{eq:ElimTruncWignerSDETrans} is numerically
integrated using a semi-implicit Milstein
algorithm~\cite{GardinerStochastic} on a spatial grid of 1024 points.
In the simulations the atoms are pumped at the cavity mode
resonance transversely with a uniform driving field $h(x)=h_0$ and with zero
axial pumping.
We set $NU
\approx 64 \hbar\omega x_0$ and $h_0^2g_0^2/\kappa\Delta_{pa}^2 \approx 0.042
\omega$. For $^{87}$Rb we have $a_s\simeq5$nm. Then, for instance, setting $x_0\simeq300$nm
and $\omega/\omega_\perp\simeq0.1$ yields $N\simeq190$.
The quantum and thermal fluctuations in the initial state are assumed
to be negligible (low temperature and large atom number, which for the given
nonlinearity translates to $N \gg 14$). The large $N$ limit also guarantees
a sufficiently fast detection rate and the validity of the adiabatic expansion
$\int(g^2(x)/\Delta_{pa}) |\psi(x)|^2 dx\ll \kappa$.

\subsection{Selective phonon excitations and their measurements}

\begin{figure}
\center{
\epsfig{file=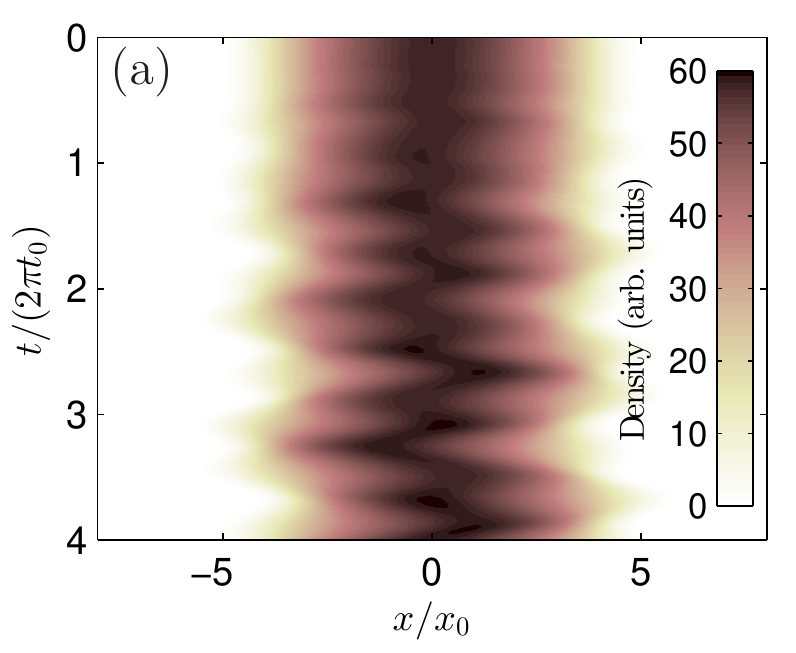}
\epsfig{file=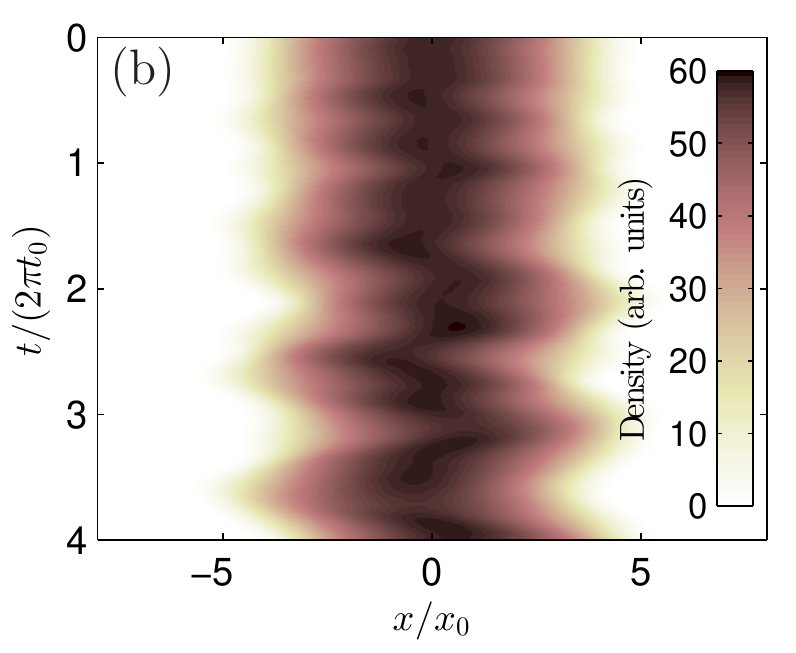}
\epsfig{file=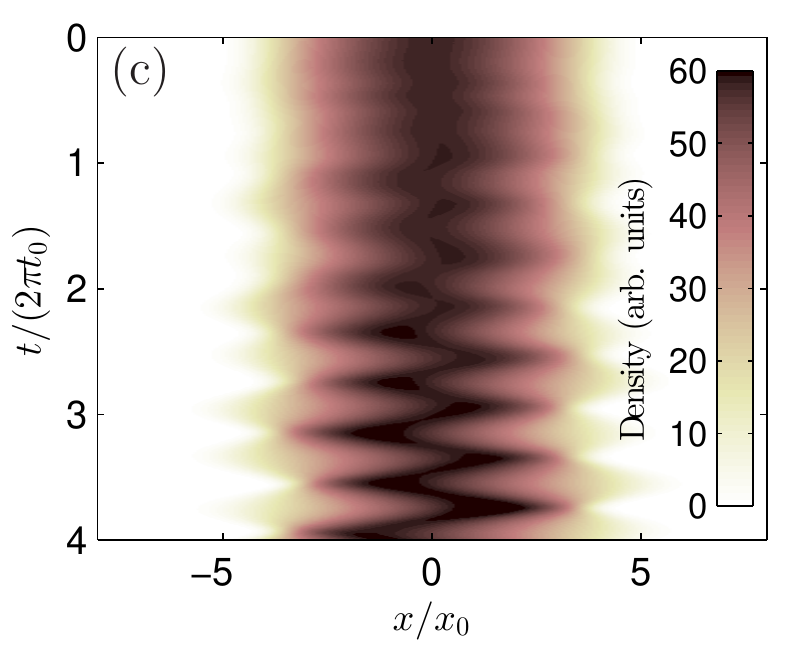}
\epsfig{file=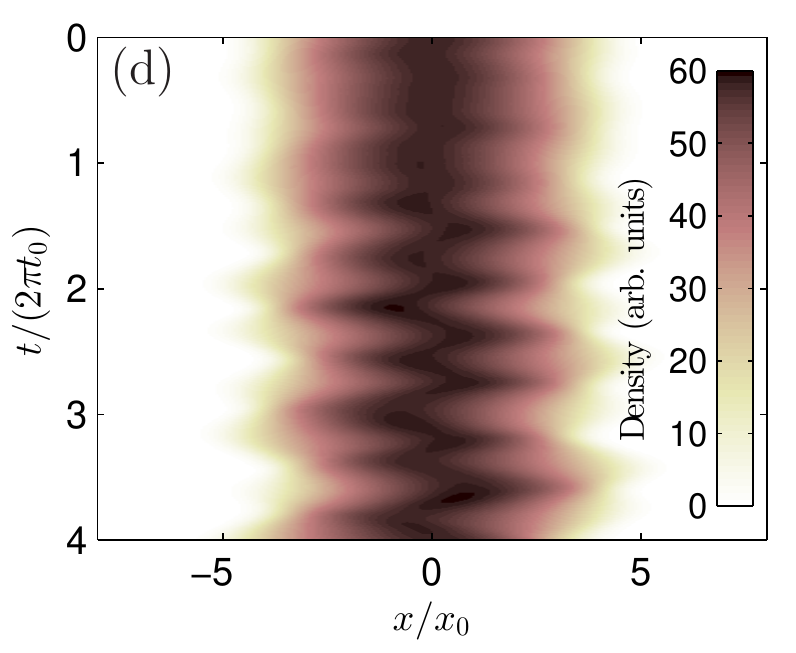}\\
\epsfig{file=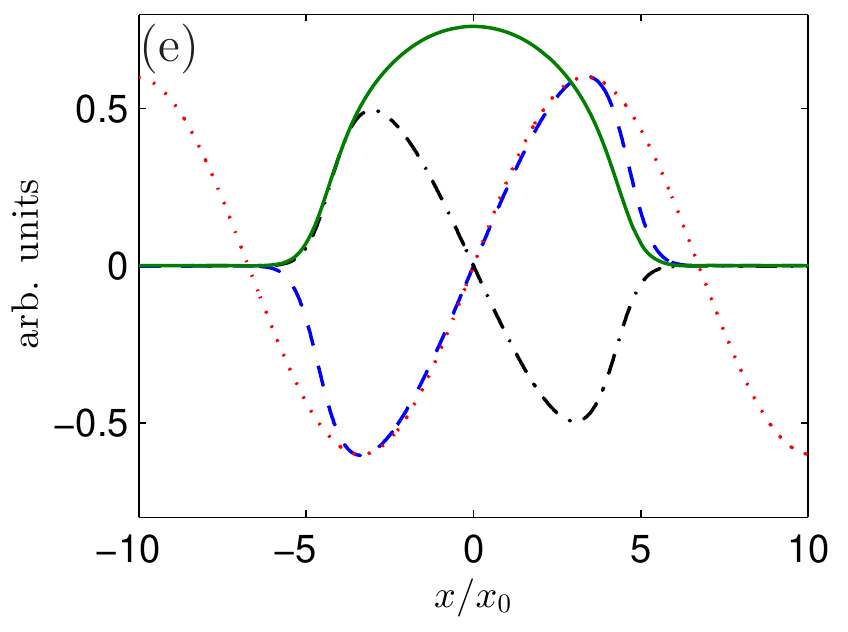,width=0.3\columnwidth}
\epsfig{file=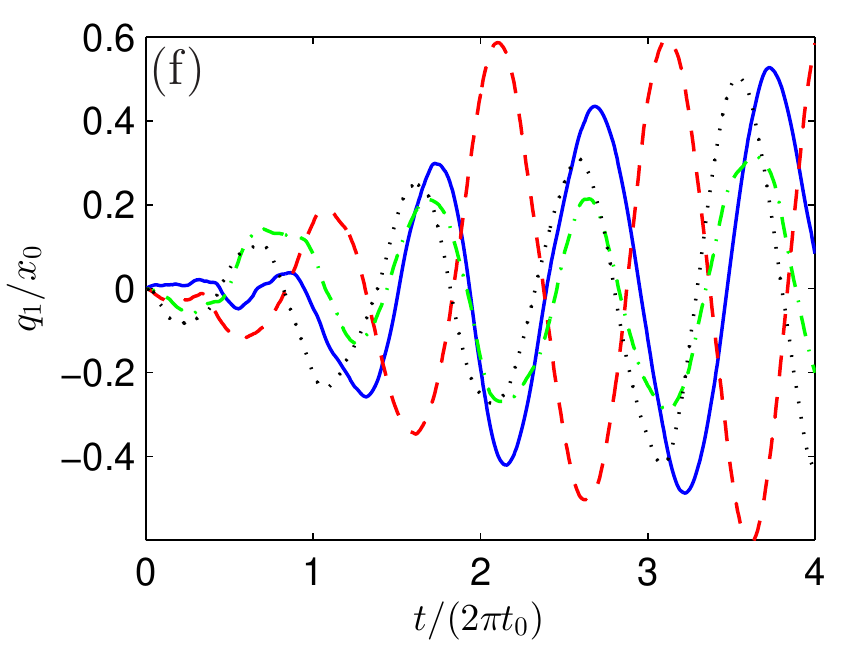,width=0.3\columnwidth}
\epsfig{file=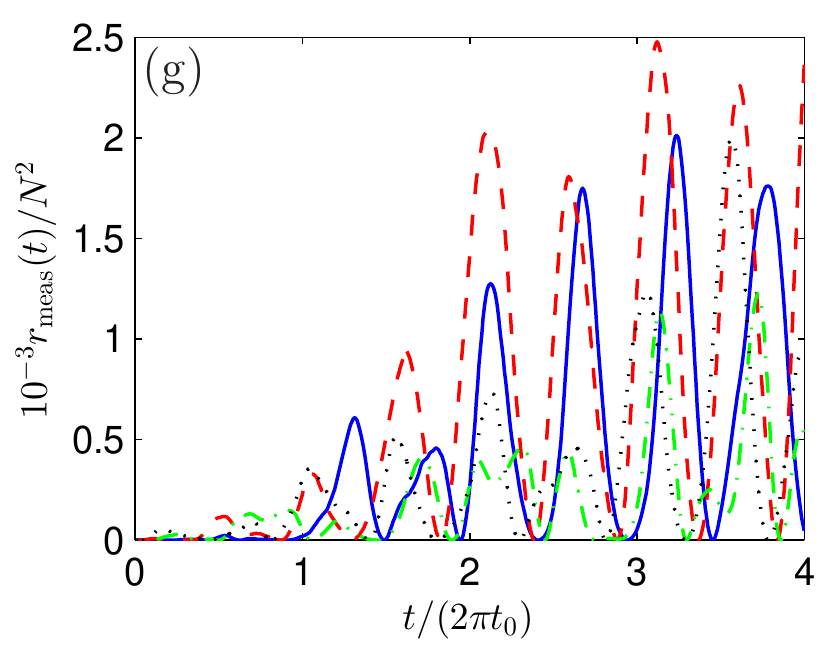,width=0.3\columnwidth}
\caption{A selective excitation of collective modes of a BEC as a result of
quantum measurement backaction. (a)-(d) The stochastic field density
$|\psi(x,t)|^2$ as a function of time for four different single realizations of
measurement trajectories in
which the Kohn mode is excited.    The transverse pump beam is applied at $t=0$
and remains at a constant strength throughout the simulation. (e) Comparison of
the shapes of the Kohn mode quasiparticle functions $u_1(x)$ (blue, dashed) and
$v_1(x)$ (black, dashed) with the cavity mode function $g(x)$ (red, dotted), and
the initial stochastic field for the condensate $\psi(x)$ (green, solid), for
the case where the overlap of the cavity mode and the Kohn mode is maximal; (e) from Ref.~\cite{measuretraj}.  The
differing quantities have been scaled into arbitrary units to enable a
comparison of their functional form. (f)-(g) The center-of-mass position $q_1(t)
= \int x |\psi(x,t)|^2 dx$ and measurement rate of photons $r_{\rm meas}(t)$
corresponding to the single realizations illustrated in (a) (blue, solid), (b)
(red, dashed), (c) (green, dot-dashed) and (d) (black, dotted).
\label{fig:MaxKohn_singleresults}}}
\end{figure}

In Fig.~\ref{fig:MaxKohn_singleresults} we show the evolution of individual
classical measurement trajectories when the dynamics conditioned on the
detection record excites the center-of-mass motion of the BEC. This induced
optomechanical coupling to a BEC corresponds to the excitation of the lowest
energy collective BEC mode (the Kohn mode).  The cavity wavelength maximizes the overlap integral $O_1$ of
\EQREF{eq:overlapintegral} in Appendix \ref{sec:decomp} for the Kohn mode, as
illustrated in Fig.~\ref{fig:MaxKohn_singleresults}(e).

In the specific numerical examples studied, the sole dynamical contribution from the cavity mode is due to
measurement backaction. The BEC is initially in the ground state and all
dynamics is therefore purely measurement induced and any differences between the
trajectories arise from the stochastic detection record. The presented
simulation results therefore also indicate a spontaneous measurement-induced
symmetry breaking in the dynamics of the atoms.

In Fig.~\ref{fig:MaxKohn_singleresults} the center-of-mass oscillations are
notable. The individual stochastic runs also considerably differ from each
other.  Collective mode amplitudes and phases also vary between realizations.  The
oscillations are revealed as pulses in the rate of photon measurements, since
the number of photons pumped into the cavity mode depends upon the time varying
overlap of the stochastic field $\psi(x,t)$ with the cavity mode $g(x)$.

\begin{figure}
\center{
\epsfig{file=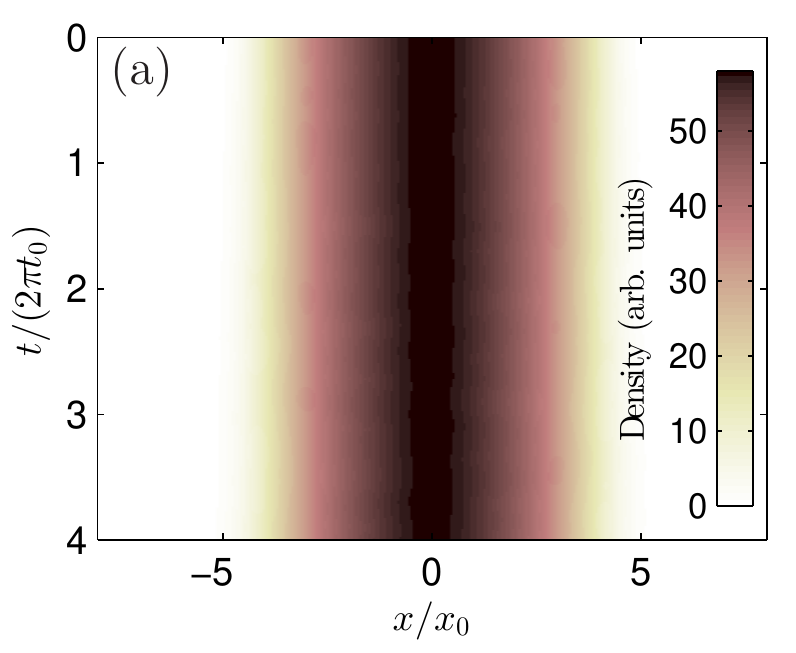,width=0.45\columnwidth}
\epsfig{file=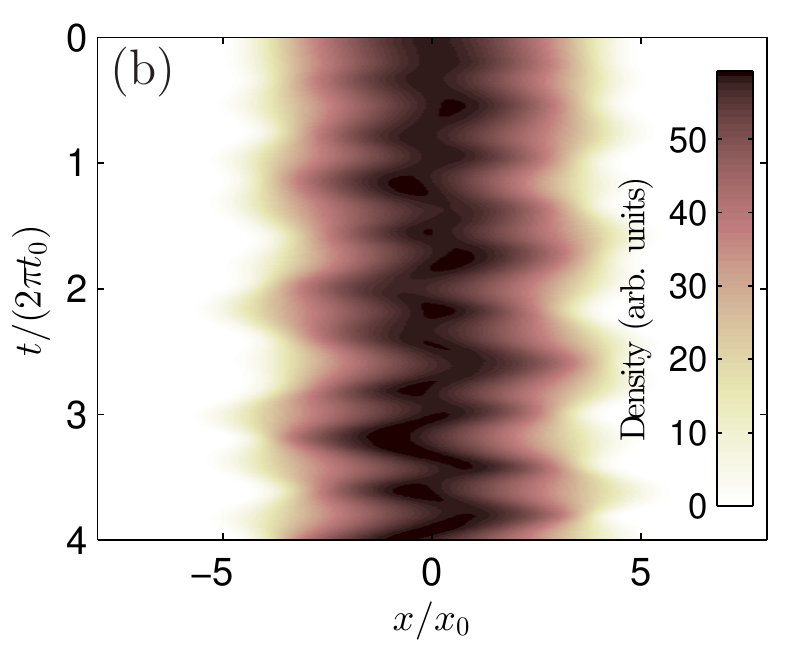,width=0.45\columnwidth}
\caption{
Measurement-induced symmetry breaking of a spatial condensate pattern and the
restored symmetry in the ensemble average over many individual realizations of
classical stochastic measurement trajectories. (a) Ensemble averaged stochastic
field density $|\psi(x,t)|^2$  of the BEC response unconditioned on any
particular measurement trajectory, formed by ensemble averaging $400$ single
trajectories. For a comparison we display it side by side with a single
trajectory that exhibits a typical
measurement-induced spatial symmetry breaking (b).
\label{fig:symmetrybreaking}}}
\end{figure}
\begin{figure}
\center{
\epsfig{file=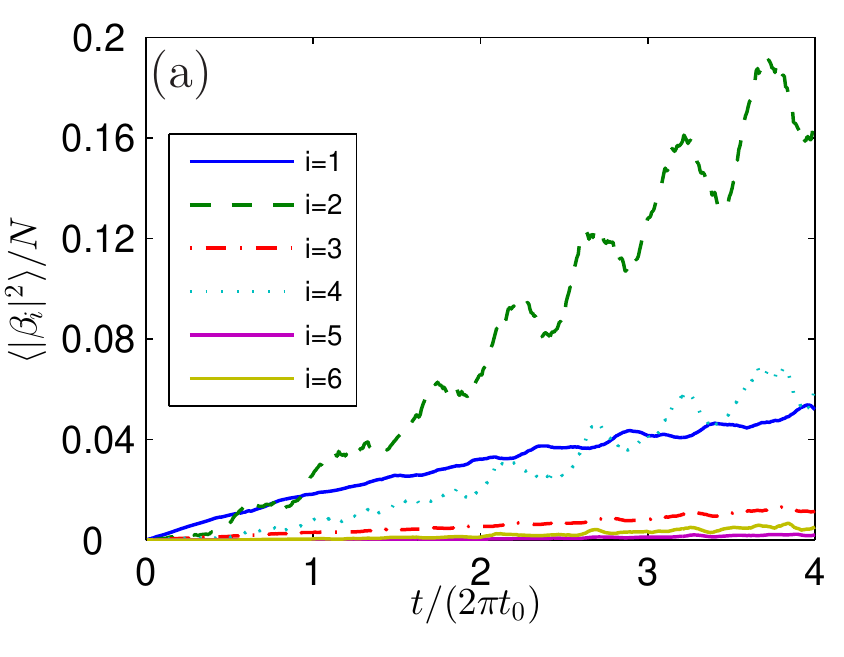,width=0.3\columnwidth}
\epsfig{file=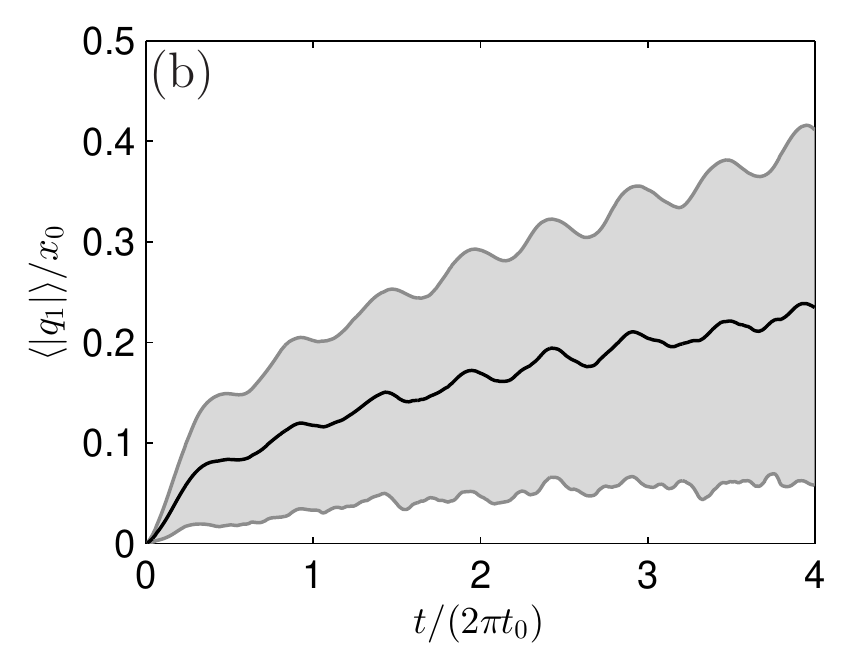,width=0.3\columnwidth}
\epsfig{file=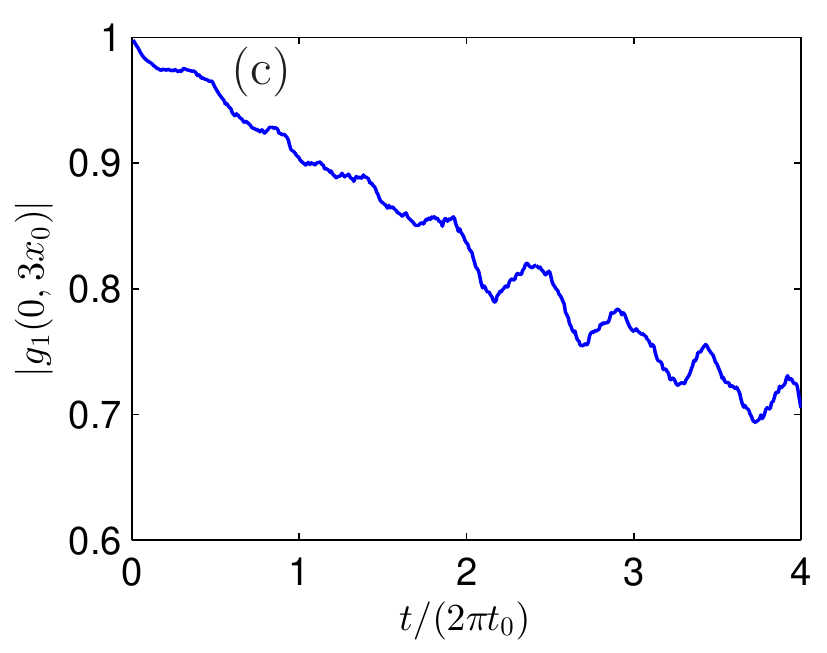,width=0.3\columnwidth}
\caption{(a) Average response of a single trajectory, decomposed into Bogoliubov
linearized collective excitations, mode occupation numbers $\langle
|\beta_i|^2\rangle$ are shown for the lowest six modes. (b) Average
center-of-mass displacement $q_1(t) = \int x |\psi(x,t)|^2 dx$ for a single
measurement conditioned trajectory. Note the absolute value must be taken since
the measurement backaction generates oscillations with a random phase for
different realizations. (c) Dissipation induced loss of coherence in the ensemble
averaged unconditioned evolution corresponding to that shown in
Fig.~\ref{fig:symmetrybreaking}(a).  The figure shows $|g_1(x,x')| \equiv
|\langle \cPsi(x)\dPsi(x')\rangle |$ between $x=0$
and $x'=3x_0$. The initial condensate is phase coherent
with $|g_1(x,x')|=1$. From Ref.~\cite{measuretraj}. \label{fig:MaxKohn_averagedresponse}}}
\end{figure}

While individual measurement trajectories show distinct effects of measurement
backaction, an ensemble average over a large number of independent trajectories
restores the initial unbroken symmetry, as shown in Fig.~\ref{fig:symmetrybreaking}.  The
ensemble averaged results reproduce the dynamics from the unconditioned FPE (\ref{simplefpe}), and such dynamics show no overall
motion of the condensate density.  The unconditioned dynamics are not trivial,
however, and the dissipation caused by the open nature of the system leads to
decoherence of the condensate [Fig.~\ref{fig:MaxKohn_averagedresponse}(c)].

In contrast to unconditioned ensemble dynamics,
Fig.~\ref{fig:MaxKohn_averagedresponse}(a)-(b) present the average response of a
single measurement conditioned trajectory.  As anticipated, from a Bogoliubov mode
decomposition it can be seen that the center-of-mass motion is largely due to an
excitation of the Kohn mode.  However, the multimode nature of the system is
illustrated by the number of other modes which are also excited.  Notably, two
even-symmetry modes respond strongly to the odd symmetry $g(x)$, a situation
made possible  once the center-of-mass motion breaks the even symmetry of
$\psi(x,t)$.

\begin{figure}
\center{
\epsfig{file=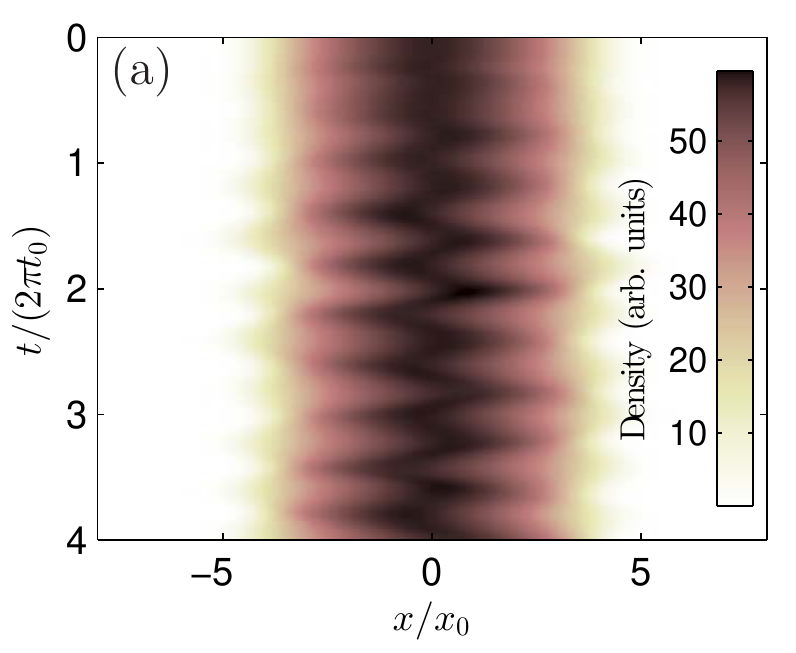,width=0.3\columnwidth}
\epsfig{file=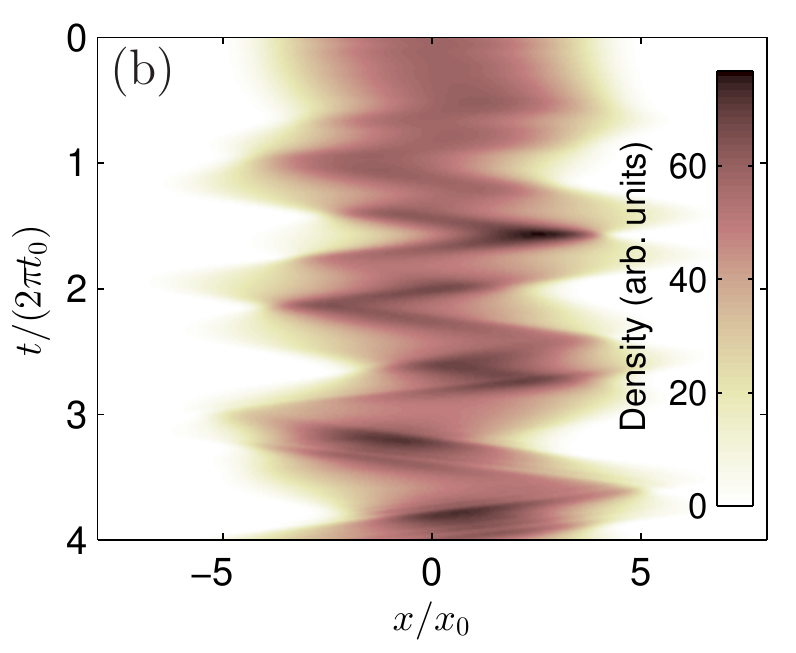,width=0.3\columnwidth}
\epsfig{file=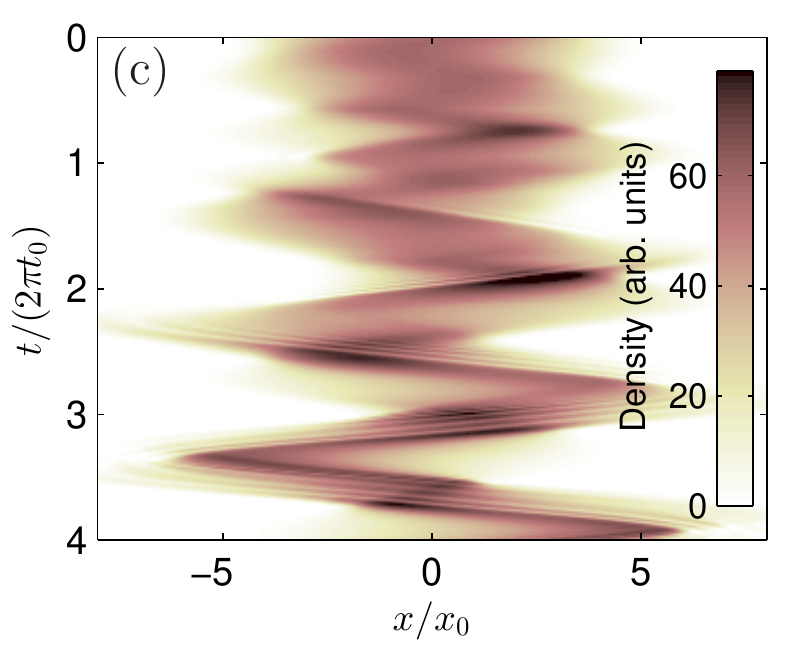,width=0.3\columnwidth}
\epsfig{file=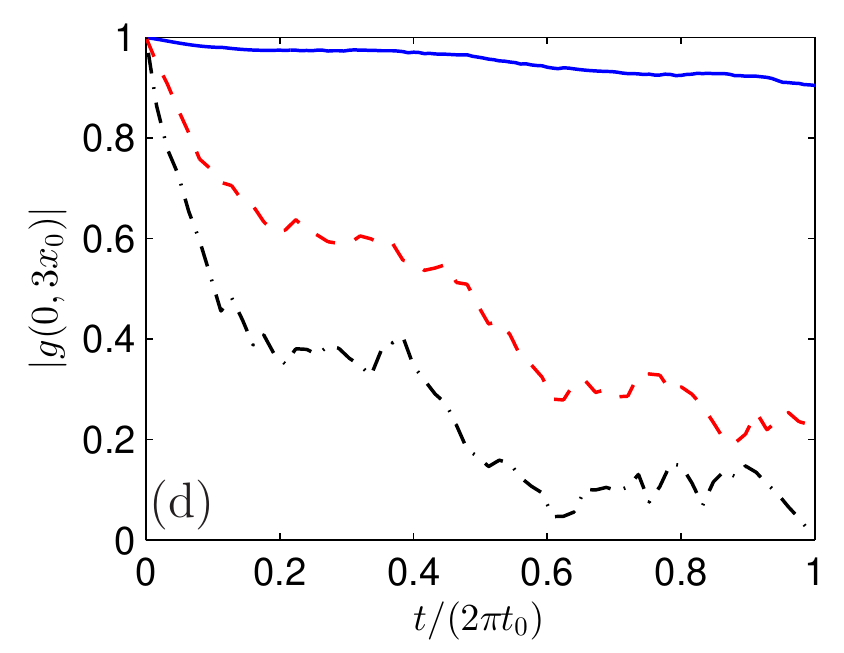,width=0.3\columnwidth}
\epsfig{file=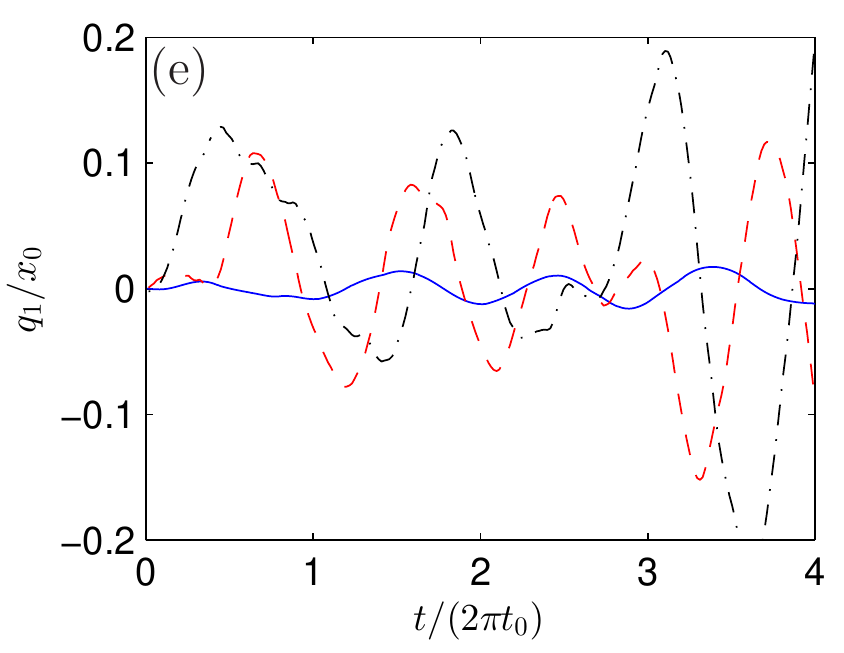,width=0.3\columnwidth}
\caption{Effect of increasing transverse pump power $h_0$ while driving the Kohn
mode.  (a) Single realization of stochastic field density $|\psi(x,t)|^2$ evolution
for pump power of $h_0 = \bar{h}_0$ with $\bar{h}_0^2g_0^2/\kappa\Delta_{pa}^2 \approx
0.042 \omega$ as was the case in Fig.~\ref{fig:MaxKohn_singleresults}. (b)
Single realization with $h_0 = 4\bar{h}_0$. (c) Single realization with $h_0 = 6\bar{h}_0$.
(d) Short time dissipation
induced loss of coherence (ensemble averaged over $200$ realizations) showing the
increase in dissipation with pump power, with $h_0 = \bar{h}_0$ (blue, solid), $h_0 =
4\bar{h}_0$ (red, dashed) and $h_0 =
6\bar{h}_0$ (black, dash-dotted).  (e) Center-of-mass displacement for the three
trajectories shown in (a)-(c), line styles as in (d).
\label{fig:PumpPower}}}
\end{figure}
Both the measurement-induced mechanical motion of single trajectories and the
decoherence of the unconditioned ensemble averaged dynamics are sensitive to the
transverse pump power, as illustrated in Fig.~\ref{fig:PumpPower}.  For larger
pump powers the induced mechanical motion becomes of larger amplitude but at a
cost of a considerable complication in the motion, as the multimode nature of the
condensate couples several intrinsic modes.    Similarly, increasing the pump
power greatly enhances the rate of decoherence of the condensate in the
unconditioned dynamics.
The induced phase decoherence of the BEC can be illustrated in terms of the density and the phase
of the atoms $\psi(x) = f(x)\exp(i\Phi(x))$, where $f^2(x)$
corresponds to the density and $\Phi(x)$ the phase. We can then obtain an alternative representation of FPE in terms of the
Wigner function $W(\{f(x),\Phi(x)\})$. The stochastic measurement part of the corresponding SDE reads
\BEQ
\left.\frac{d f(x)}{dt}\right|_{\mathrm{meas.}} = 0, \quad
\left.d\Phi(x)\right|_{\mathrm{meas.}} =
\sqrt{\frac{2}{\kappa}}\frac{g(x)h(x)}{\Delta_{pa}(x)}dW\,.
\label{eq:dPhieqn}
\EEQ
The measurement has a direct effect on the phase evolution of the atoms in such a
way that the phase profile can considerably fluctuate between different individual
runs. The spatially non-trivial dependence of these fluctuations results from the
cavity field as well as from the transverse pump profile and detuning. No similar
direct effect on the density profile exists and the density variation is a
consequence of the phase dynamics. Averaging over many trajectories yields the
loss of coherence, as also illustrated in Fig.~\ref{fig:PumpPower}.

The spatially non-trivial dependence of fluctuations introduces a sensitivity
of the measurements to the position of the atom cloud. In the limit of weak interactions,
the atom cloud in the ground state would be approximated by the harmonic oscillator ground state
wavefunction close to a potential minimum. In this limit the atoms would undergo zero point motion.

\begin{figure}
\center{
\epsfig{file=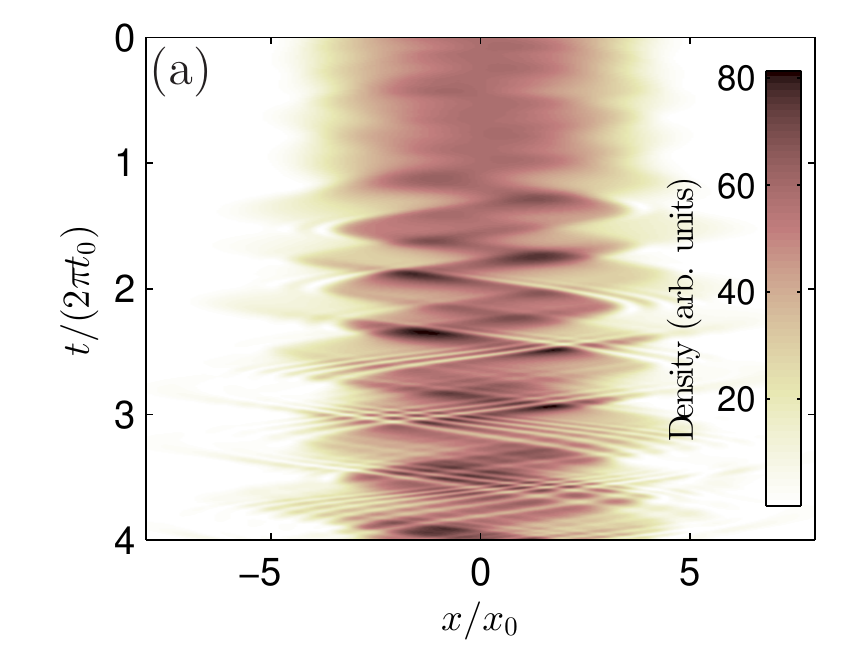,width=0.45\columnwidth}
\epsfig{file=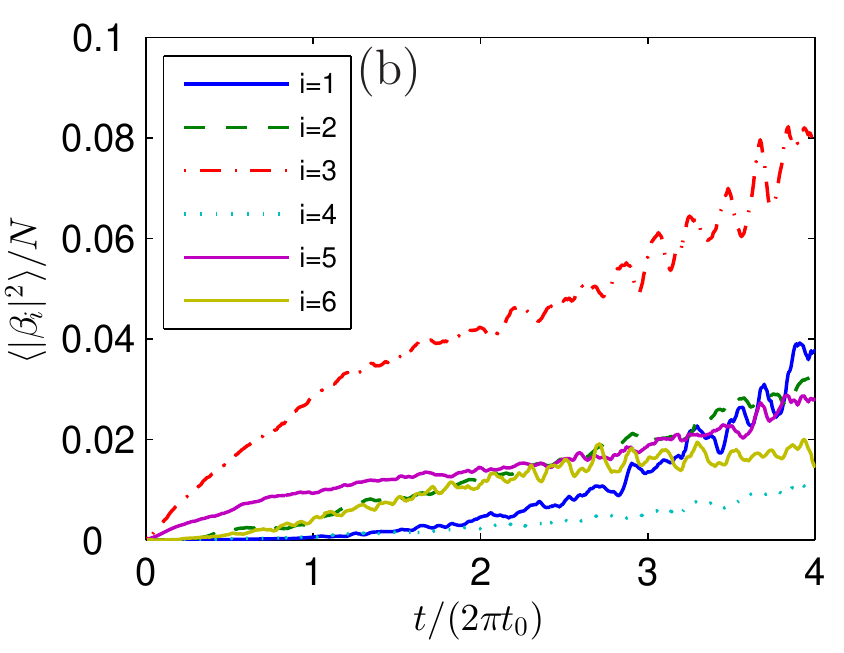,width=0.45\columnwidth}
\caption{Response for a cavity wavelength of $k=1.03x_0^{-1}$, such that the
overlap integral $O_3$ of \EQREF{eq:overlapintegral} provides maximal coupling
to the third Bogoliubov mode. (a) Density response of the stochastic field $|\psi(x,t)|^2$ for
a single stochastic realization of a measurement record.  (b) Average single
trajectory Bogoliubov mode populations, averaged over $400$ realizations. (b) from Ref.~\cite{measuretraj}.
\label{fig:MaxKohn_VaryLambda_vl2gk3}}}
\end{figure}

Owing to the multimode nature, the BEC exhibits several collective excitation modes that can be selectively driven.
By altering the ratio of cavity wavelength to trap length we can target
different Bogoliubov collective modes.  The next Bogoliubov mode with an odd symmetry is the one with the third lowest energy.
Figure~\ref{fig:MaxKohn_VaryLambda_vl2gk3} shows
the results with cavity wavenumber $k=1.03x_0^{-1}$, chosen to predominantly
excite this mode.

\begin{figure}
\center{
\epsfig{file=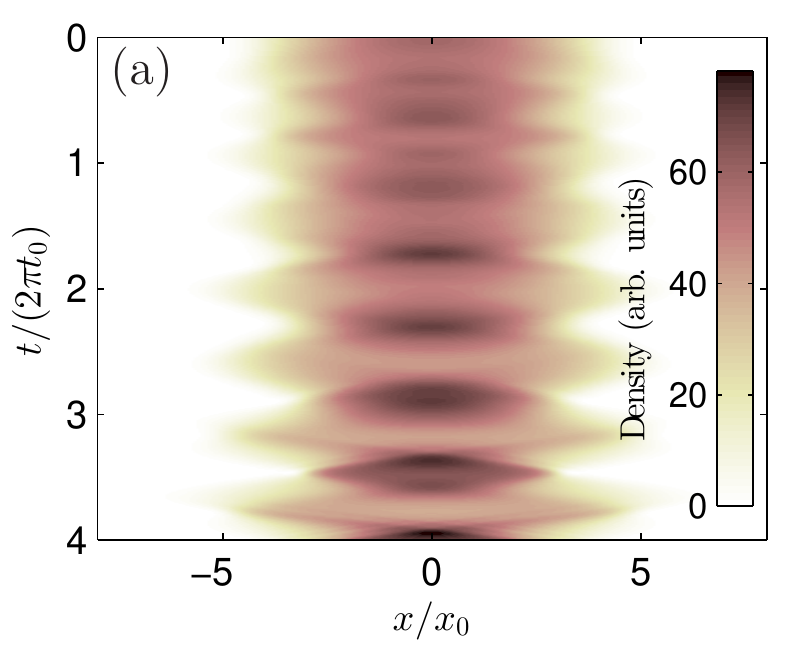,width=0.45\columnwidth}
\epsfig{file=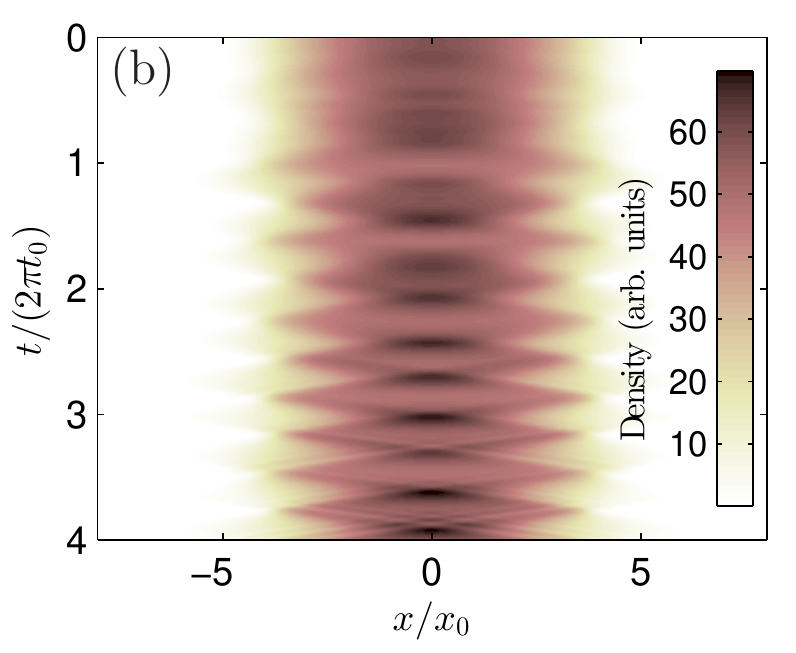,width=0.45\columnwidth} \\
\epsfig{file=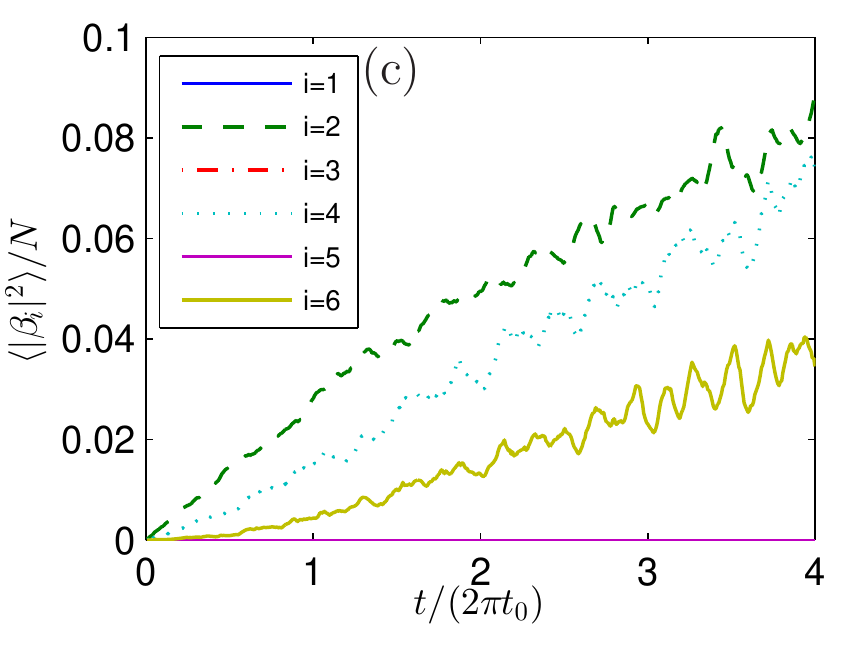,width=0.45\columnwidth}
\epsfig{file=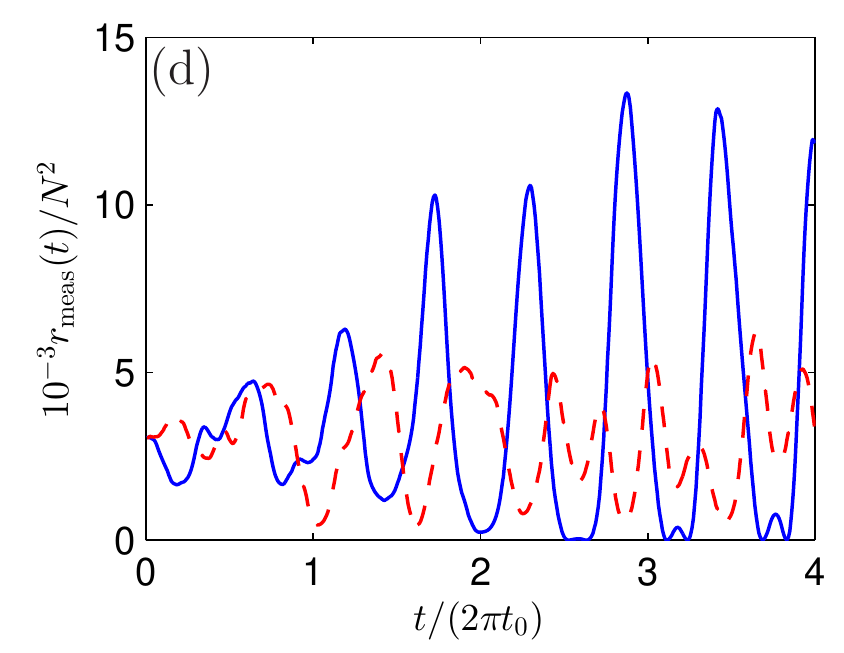,width=0.45\columnwidth}
\caption{Response of the condensate to measurement backaction when the cavity
mode coupling $g(x)$ is tailored to give a maximum overlap with the breathing
mode, centering the trap potential on a cavity antinode such that $g(x) =
g_0\cos(kx)$. (a)-(b) Two single realizations of the condensate stochastic field density
$|\psi(x,t)|^2$.  (c) Bogoliubov mode populations, averaged over $400$ realizations (from Ref.~\cite{measuretraj}).
(d) Measurement rate $r_{\rm meas}(t)$ corresponding to the density evolutions
shown in (a) (blue, solid) and (b) (red, dashed).
\label{fig:MaxBreathe}}}
\end{figure}

By further changing the symmetry of $g(x)$ with respect to the trap center, that is by taking $g(x) =
g_0 \cos(kx)$, we can selectively target modes of even symmetry.  Figure~\ref{fig:MaxBreathe}
shows the targeted excitation of the lowest even mode, the breathing mode.
Contrary to the case where odd modes are excited, the motion of an even symmetry
mode does not change the symmetry properties of $\psi(x,t)$ with respect to the
trap center, and consequently no odd modes are excited by this choice of
$g(x)$.  A possible practical problem for such cases may arise when the measured
photon rate is considered.  For even symmetry $g(x)$ the initial static
condensate already generates a significant photon measurement rate, and so
detection of oscillations in the photon measurement rate due to excitation of
collective modes may be more difficult than in the odd symmetry case.  However,
in the case of the breathing mode, and for our choice of parameters, the signals
in the photon measurement rate are clear.

\subsection{Phonon detection}

In the previous section we discussed the selective excitation of phonons and the
optical signatures of phonon modes when strong phonon excitations were present. It
would be particularly promising to employ cavity-BEC systems for the detection of
weakly excited phonon modes and even as potential single phonon detectors. Here we
show that by counting photons outside the cavity that are almost solely scattered
by the phonons may accurately detect weak phonon excitations.

In order to analyze phonon modes quantum-mechanically, we take the Bogoliubov
expansion \eqref{field} and assume that only the BEC mode has a large occupation.
Consequently, we replace $\h \beta_0 \sim \sqrt{N_0}$, where $N_0$ denotes the
condensate particle number. For simplicity, in the following we also assume that
the mode functions $\psi_0(x)$, $u_j(x)$, and $v_j(x)$ are real.

\begin{figure}
\center{
\epsfig{file=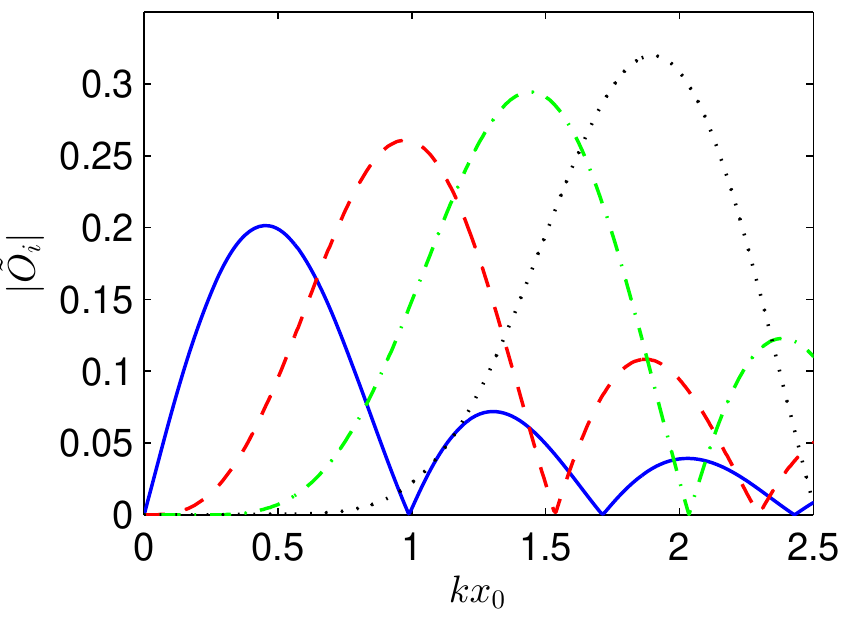,width=0.48
\columnwidth}
\caption{Overlap integral relevant to phonon detection in a transversely
pumped cavity $\tilde{O}_i = 1/(g_0h_0)\int
g(x)h(x)\left[u_i(x)-v_i(x)\right]\psi_0(x) dx$, as a function of the
cavity wavenumber $k$, where $g(x) = g_0\sin(kx)$ and $h(x) = h_0$ is
uniform.  The condensate is centered at $x=0$, a node of the cavity
light amplitude.  We show such overlaps for the four lowest energy odd
Bogoliubov modes $i=1$ (blue, solid), $i=3$ (red, dashed), $i=5$ (green,
dot-dashed), and $i=7$ (black, dotted).  Note that the symmetry of the
system implies $\tilde{O}_i = 0$ for even numbered modes, and the
overlap with the condensate density of Eq.~(\ref{eq:condensateoverlap}) also vanishes in this system.
\label{fig:BdGOverlap}}}
\end{figure}

We further restrict the discussion to considering the detection of weak phonon
excitations in a setup where the cavity is transversely pumped, such that the
observable is $\h Y$. However, the following treatment can be adapted to the
cavity pumping case by a straightforward change to the operator $\h X$. The light
intensity measured outside the cavity consists of the coherently scattered part
and the fluctuations $I_{\rm tot} = I_{\rm coh} + I_{\rm fl}$. The coherent part $
\propto |\< \h a\>|^2 $ then depends on
\beq
\<\h a \> = {1\over \kappa}\< \h Y \> \simeq {1\over \kappa}\int dx\, {g(x) h(x)\over \Delta_{pa}} \big\{ N_0 \psi_0^2(x)+ \sum_j [u_j^2(x) n_j + v_j^2(x) (n_j +1) ]\big\}\,.
\label{eq:amplphd}
\eeq
The last expression with vanishing phonon excitations ($n_j=0$) represents the
condensate quantum depletion. For measurements of the light intensity
fluctuations outside the cavity, we need to consider the observable
\beq
\<\h a^\dagger \h a\> - |\< \h a\>|^2 = {1\over \kappa^2} (\Delta \h Y)^2,
\eeq
where
\beq
(\Delta \h Y)^2 = \< \h Y^2\> - \< \h Y \>^2\simeq N_0 \sum_j \left\{ \int dx\, {g(x) h(x)\over \Delta_{pa}} [u_j(x)-v_j(x)]  \psi_0  (x)    \right\}^2 (2n_j+1)\,.
\label{eq:lightfluct}
\eeq
Here we have only kept the leading order contribution (for large $N_0$). The exact
nature of the scattered intensity now depends on the choice of $g(x)h(x)$.  For a
system similar to that described in the previous section, with a uniform pump and
a condensate trapped at a node of $g(x)$, then the symmetry of the system ensures
that \EQREF{eq:amplphd} vanishes.  Equation~(\ref{eq:lightfluct}) is then the
dominant term in the scattered light, and measurements of a given phonon mode can
be made by maximizing the corresponding integral term.
Figure~\ref{fig:BdGOverlap} shows such overlaps for the odd numbered modes, while
even mode overlaps vanish. The measurement rate of photons corresponding to \EQREF{eq:lightfluct} does not need to be small and can also be considerably varied. For instance, for the parameters
given in section 3.1, for $N = 1000$ we obtain the counting
rate of about $10\omega$.

In general, the pump mode shape $h(x)$ can be tailored and the symmetry of the
cavity mode can be altered, enabling phonon measurement either through maximizing
\EQREF{eq:lightfluct}, as in Fig.~\ref{fig:BdGOverlap}, or by a minimising of
\EQREF{eq:lightfluct} such that the higher order terms dominate.  The latter case is
analogous to trying to influence \EQREF{eq:amplphd} such that the second term is
comparable to the first term, and requires the integrals
\beq
\int dx\, {g(x) h(x)\over \Delta_{pa}} u_j^2(x), \quad \int dx\, {g(x) h(x)\over \Delta_{pa}} v_j^2(x)\,,
\eeq
to be maximized for some $j$ whilst simultaneously minimizing the overlap integral
\beq
\int dx\, {g(x) h(x)\over \Delta_{pa}} \psi_0^2(x)\,.
\label{eq:condensateoverlap}
\eeq

The detection of quantum correlations of atoms confined in an optical cavity has
previously been proposed, e.g., as an unambiguous signature of whether the atoms
are in a Mott-insulator or a superfluid state \cite{PhysRevLett.98.100402}. More
generally, light can act as a sensitive diagnostic tool of ultracold atoms in
periodic lattice systems, see
e.g.~\cite{ruostekoski:170404,PhysRevA.84.053608,PhysRevA.80.043404,PhysRevA.81.013404,
PhysRevA.81.013415,AFMdetection}. In addition, BEC-cavity systems
have previously been considered, e.g., in weak force sensing applications~\cite{odell1,Samoylova:15}.

Using the BEC-cavity system as a sensitive phonon detector could measure the
statistical properties of phonons and therefore act as an accurate BEC thermometer
or detect quantum correlations. The backaction of the quantum measurement process
could be employed in preparation of complex quantum states of phonons. The phonon detector, based on
the BEC-cavity system can also be considered as a counterpart to the idea to use a BEC as a single photon detector in a cavity~\cite{Horak_singlephoton,Horak_beccavity}.

\section{Concluding remarks}

Coupling interacting ultracold atoms to optical cavity fields provides a
promising system to study several many-body realizations of quantum optical
technologies, such as quantum state preparation and the effect of quantum
measurement. Here we have investigated individual runs of continuously observed
BECs in a cavity where the dynamics is conditioned on the measurement record.
The method is based on classical measurement trajectories~\cite{measuretraj}
that can numerically incorporate large many-atom systems with over a thousand
spatial grid points by approximating the atom-cavity dynamics. Tailoring of the
BEC-cavity system allows for the selective coupling of light to collective
excitations of the condensate. In the weak excitation limit this could be
applied to the development of a sensitive phonon detector.

In the ultracold many-atom context the effect of quantum measurement and
continuous monitoring of the dynamics have attracted considerable theoretical
interest in the measurement schemes of atom
counting~\cite{Javanainen1996a,Cirac1996a,Castin1997a}, in the photon
counting~\cite{RUO98,Ruostekoski1997a,javanainen10,Elliott_QM} and homodyne
measurements~\cite{Corney1998a} of scattered light, in dispersive
imaging~\cite{Dalvit2002a,Szigeti2009a}, and in an interferometric
context~\cite{Lee2012a}. Semiclassical~\cite{Niedenzu2013a} and static discrete
approximations~\cite{Mekhov2009a} have been considered for larger atom clouds in
cavities. Feedback-control mechanisms~\cite{Wiseman2010a} have also been
employed in ultracold atom setups in approximate approaches~\cite{Hush2013a}. In
fact, simulations of the emergence of a relative phase between two BECs in a
continuous quantum measurement process even when the BECs initially have no
relative phase
information~\cite{Javanainen1996a,Cirac1996a,Castin1997a,Ruostekoski1997a}
represent a measurement-induced spontaneous symmetry breaking analogous to that
shown in our simulations (see Fig.~\ref{fig:symmetrybreaking}). In the phase
measurements the phase is initially entirely random, but a continuous detection
process eventually establishes a well-defined phase. Since in each simulation
run this value emerges randomly, ensemble-averaging over many realizations
results in a flat phase distribution $[0,2\pi[$ and a fully restored symmetry,
analogously to the restored symmetry of the density profile of
Fig.~\ref{fig:symmetrybreaking}(b).

\acknowledgments{Acknowledgments}

This work was supported financially by the EPSRC.


\authorcontributions{Author Contributions}
Both authors contributed to the research and writing of the manuscript. The numerical simulations were performed by MDL.


\conflictofinterests{Conflicts of Interest}

The authors declare no conflict of interest.

\appendix
\section{Analyzing the collective mode response}
\label{sec:decomp}

The excitation of different collective modes could be estimated in the limit of weak excitations
by decomposing the BEC motion into the linearized Bogoliubov modes. Our treatment is
related to the cavity cooling analysis in~\cite{Gardiner2001a}. We use a classical version of the expansion \EQREF{field},
\beq
\psi(x,t) = e^{-i\mu
t/\hbar}\bigg\{\beta_0(t)\psi_0(x) +\sum_{i\neq 0}\left[\beta_i(t)u_i(x)-\beta_i^*(t)v_i^*(x)\right]\bigg\},
\eeq
where the quasiparticle modes $u_i(x)$ and $v_i(x)$ are the solutions to the Bogoliubov equations and the mode amplitudes $\beta_j$ are now treated classically. The expansion may be inverted to give the mode amplitudes as a function of the stochastic field from the classical measurement trajectories
\BEQ
\beta_i = \int \mbox{d}x \left[u^*_i(x)\psi(x)e^{i\mu t/\hbar}+v_i^*(x)\psi^*(x)e^{-i\mu
t/\hbar}\right]\,.
\EEQ

The measurement of light intensity leaking out of the cavity represents the specific measurement of the atomic operator $\hat{Y}$, defined in \EQREF{eq:Ydefn}. We may now expand $\hat{Y}$ into the collective modes of the BEC inside the cavity
\beq
\av{\hat{Y}} = \frac{h_0}{\Delta_{pa}}\bigg\{|\beta_0|^2\int g(x)|\psi_0(x)|^2 dx
+\sum_{i\neq 0}
\beta_0^*\int g(x)\psi^*_0(x)\left[\beta_iu_i(x)-\beta_i^*v_i(x)\right] dx\bigg\}\,, \label{eq:YdecompintoBdG}
\eeq
where we have assumed $|\beta_0|^2 \gg |\beta_j|^2$ for $j\neq 0$. We can therefore try to maximize the
measurement backaction on specific collective mode $i$ by
maximizing the corresponding overlap integral
\BEQ
O_i = \int g(x) \psi_0^*(x)\left[u_i(x)-v_i(x)\right] dx\,.
\label{eq:overlapintegral}
\EEQ


\bibliographystyle{mdpi}

\begin{thebibliography}{----}
\providecommand{\natexlab}[1]{#1}

\bibitem[Carmichael(2007)]{CarmichaelVol2}
Carmichael, H.
\newblock {\em Statistical Methods in Quantum Optics Vol 2}; Springer: Berlin, Germany 2007.

\bibitem[Brennecke \em{et~al.}(2007)Brennecke, Donner, Ritter, Bourdel, Kohl,
and Esslinger]{Brennecke2007a}
Brennecke, F.; Donner, T.; Ritter, S.; Bourdel, T.; Kohl, M.; Esslinger, T.
\newblock Cavity {QED} with a~Bose-Einstein condensate.
\newblock {\em Nature} {\bf 2007}, {\em 450},~268--271.

\bibitem[Colombe \em{et~al.}(2007)Colombe, Steinmetz, Dubois, Linke, Hunger,
and Reichel]{Colombe2007a}
Colombe, Y.; Steinmetz, T.; Dubois, G.; Linke, F.; Hunger, D.; Reichel, J.
\newblock Strong atom-field coupling for Bose-Einstein condensates in an
optical cavity on a chip.
\newblock {\em Nature} {\bf 2007}, {\em 450},~272--276.

\bibitem[Murch \em{et~al.}(2008)Murch, Moore, Gupta, and
Stamper-Kurn]{Murch2008a}
Murch, K.W.; Moore, K.L.; Gupta, S.; Stamper-Kurn, D.M.
\newblock Observation of quantum-measurement backaction with an ultracold
atomic gas.
\newblock {\em Nat. Phys.} {\bf 2008}, {\em 4},~561--564.

\bibitem[Brahms \em{et~al.}(2012)Brahms, Botter, Schreppler, Brooks, and
Stamper-Kurn]{Brahms2012a}
Brahms, N.; Botter, T.; Schreppler, S.; Brooks, D.W.C.; Stamper-Kurn, D.M.
\newblock Optical Detection of the Quantization of Collective Atomic Motion.
\newblock {\em Phys. Rev. Lett.} {\bf 2012}, doi:10.1103/PhysRevLett.108.133601.

\bibitem[Botter \em{et~al.}(2013)Botter, Brooks, Schreppler, Brahms, and
Stamper-Kurn]{Botter2013a}
Botter, T.; Brooks, D.W.C.; Schreppler, S.; Brahms, N.; Stamper-Kurn, D.M.
\newblock Optical Readout of the Quantum Collective Motion of an Array of
Atomic Ensembles.
\newblock {\em Phys. Rev. Lett.} {\bf 2013}, doi:10.1103/PhysRevLett.110.153001.

\bibitem[Schmidt \em{et~al.}({2014})Schmidt, Tomczyk, Slama, and
Zimmermann]{Zimmermann}
Schmidt, D.; Tomczyk, H.; Slama, S.; Zimmermann, C.
\newblock {Dynamical Instability of a Bose-Einstein Condensate in an Optical
Ring Resonator}.
\newblock {\em {Phys. Rev. Lett.}} {\bf {2014}}, doi:10.1103/PhysRevLett.112.115302.

\bibitem[Kessler \em{et~al.}({2014})Kessler, Klinder, Wolke, and
Hemmerich]{Hemmerich}
Kessler, H.; Klinder, J.; Wolke, M.; Hemmerich, A.
\newblock {Steering Matter Wave Superradiance~with an Ultranarrow-Band Optical
Cavity}.
\newblock {\em {Phys. Rev. Lett.}} {\bf {2014}}, doi:10.1103/PhysRevLett.113.070404.

\bibitem[Ritsch \em{et~al.}(2013)Ritsch, Domokos, Brennecke, and
Esslinger]{Ritsch2013a}
Ritsch, H.; Domokos, P.; Brennecke, F.; Esslinger, T.
\newblock Cold atoms in cavity-generated dynamical optical potentials.
\newblock {\em Rev. Mod. Phys.} {\bf 2013}, {\em 85},~553--601.

\bibitem[Brennecke \em{et~al.}(2008)Brennecke, Ritter, Donner, and
Esslinger]{Brennecke2008a}
Brennecke, F.; Ritter, S.; Donner, T.; Esslinger, T.
\newblock Cavity Optomechanics with a Bose-Einstein Condensate.
\newblock {\em Science} {\bf 2008}, {\em 322},~235 --238.

\bibitem[Tian and Carmichael(1992)]{Tian1992a}
Tian, L.; Carmichael, H.J.
\newblock Quantum trajectory simulations of two-state behavior in an optical
cavity containing one atom.
\newblock {\em Phys. Rev. A} {\bf 1992}, {\em 46},~R6801--R6804.

\bibitem[Dalibard \em{et~al.}(1992)Dalibard, Castin, and
M\o{}lmer]{Dalibard1992a}
Dalibard, J.; Castin, Y.; M\o{}lmer, K.
\newblock Wave-function approach to dissipative processes in quantum optics.
\newblock {\em Phys. Rev. Lett.} {\bf 1992}, {\em 68},~580--583.

\bibitem[Dum \em{et~al.}(1992)Dum, Zoller, and Ritsch]{Dum1992a}
Dum, R.; Zoller, P.; Ritsch, H.
\newblock Monte Carlo simulation of the atomic master equation for spontaneous
emission.
\newblock {\em Phys. Rev. A} {\bf 1992}, {\em 45},~4879--4887.

\bibitem[Leroux \em{et~al.}(2010)Leroux, Schleier-Smith, and
Vuleti\ifmmode~\acute{c}\else \'{c}\fi{}]{spinsqueezing_vuletic}
Leroux, I.D.; Schleier-Smith, M.H.; Vuleti\ifmmode~\acute{c}\else \'{c}\fi{},
V.
\newblock Implementation of Cavity Squeezing of a Collective Atomic Spin.
\newblock {\em Phys. Rev. Lett.} {\bf 2010}, doi:10.1103/PhysRevLett.104.073602.

\bibitem[Lee and Ruostekoski(2014)]{measuretraj}
Lee, M.D.; Ruostekoski, J.
\newblock Classical stochastic measurement trajectories: Bosonic atomic gases
in an optical cavity and quantum measurement backaction.
\newblock {\em Phys. Rev. A} {\bf 2014}, doi:10.1103/PhysRevA.90.023628.

\bibitem[Javanainen and Ruostekoski(2013)]{Javanainen2013a}
Javanainen, J.; Ruostekoski, J.
\newblock Emergent classicality in continuous quantum measurements.
\newblock {\em New J. Phys.} {\bf 2013}, doi;10.1088/1367-2630/15/1/013005.

\bibitem[Gardiner \em{et~al.}(2001)Gardiner, Gheri, and Zoller]{Gardiner2001a}
Gardiner, S.A.; Gheri, K.M.; Zoller, P.
\newblock Cavity-assisted quasiparticle damping in a Bose-Einstein condensate.
\newblock {\em Phys. Rev. A} {\bf 2001}, doi:10.1103/PhysRevA.63.051603.

\bibitem[Nagy \em{et~al.}(2008)Nagy, Szirmai, and Domokos]{Nagy2008a}
Nagy, D.; Szirmai, G.; Domokos, P.
\newblock Self-organization of a Bose-Einstein condensate in an optical cavity.
\newblock {\em Eur. Phys. J. D} {\bf 2008}, {\em 48},~127--137.

\bibitem[Szirmai \em{et~al.}(2009)Szirmai, Nagy, and Domokos]{Szirmai2009a}
Szirmai, G.; Nagy, D.; Domokos, P.
\newblock Excess Noise Depletion of a Bose-Einstein Condensate in an Optical
Cavity.
\newblock {\em Phys. Rev. Lett.} {\bf 2009}, doi:10.1103/PhysRevLett.102.080401.

\bibitem[Venkatesh and O'Dell(2013)]{odell2}
Venkatesh, B.P.; O'Dell, D.H.J.
\newblock Bloch oscillations of cold atoms in a cavity: Effects of quantum
noise.
\newblock {\em Phys. Rev. A} {\bf 2013}, doi:10.1103/PhysRevA.88.013848.

\bibitem[Jaynes and Cummings(1963)]{Jaynes1963a}
Jaynes, E.; Cummings, F.W.
\newblock Comparison of quantum and semiclassical radiation theories with
application to the beam maser.
\newblock {\em Proc. {IEEE}} {\bf 1963}, {\em 51},~89--109.

\bibitem[Walls and Milburn(1994)]{WallsMilburn}
Walls, D.F.; Milburn, G.J.
\newblock {\em Quantum Optics}; 2nd ed.; Springer: Berlin, Germany,   1994.

\bibitem[Maschler \em{et~al.}(2008)Maschler, Mekhov, and Ritsch]{Maschler2008a}
Maschler, C.; Mekhov, I.B.; Ritsch, H.
\newblock Ultracold atoms in optical lattices generated by quantized light
fields.
\newblock {\em Eur. Phys. J. D} {\bf 2008}, {\em 46},~545--560.

\bibitem[Carmichael(1993)]{Carmichael1993a}
Carmichael, H.
\newblock {\em An Open Systems Approach to Quantum Optics}; Springer-Verlag, Berlin, Germany,
1993.

\bibitem[Gardiner and Zoller(2004)]{QuantumNoise}
Gardiner, C.; Zoller, P.
\newblock {\em Quantum Noise}; Springer, Berlin, Germany, 2004.

\bibitem[Olshanii(1998)]{olshanii_98}
Olshanii, M.
\newblock Atomic Scattering in the Presence of an External Confinement and a
Gas of Impenetrable Bosons.
\newblock {\em Phys. Rev. Lett.} {\bf 1998}, {\em 81},~938--941.

\bibitem[Ruostekoski and Isella(2005)]{RUO05}
Ruostekoski, J.; Isella, L.
\newblock Dissipative Quantum Dynamics of Bosonic Atoms in a Shallow 1D Optical
Lattice.
\newblock {\em Phys. Rev. Lett.} {\bf 2005}, doi:10.1103/PhysRevLett.95.110403.

\bibitem[Carmichael(1999)]{CarmichaelVol1}
Carmichael, H.
\newblock {\em Statistical Methods in Quantum Optics Vol 1}; Springer, Berlin, Germany, 1999.

\bibitem[Drummond and Hardman(1993)]{drummond93}
Drummond, P.D.; Hardman, A.D.
\newblock Simulation of Quantum Effects in Raman-Active Waveguides.
\newblock {\em EPL (Europhys. Lett.)} {\bf 1993}, doi:10.1209/0295-5075/21/3/005.

\bibitem[Steel \em{et~al.}(1998)Steel, Olsen, Plimak, Drummond, Tan, Collett,
Walls, and Graham]{Steel1998a}
Steel, M.J.; Olsen, M.K.; Plimak, L.I.; Drummond, P.D.; Tan, S.M.; Collett,
M.J.; Walls, D.F.; Graham, R.
\newblock Dynamical quantum noise in trapped Bose-Einstein condensates.
\newblock {\em Phys. Rev. A} {\bf 1998}, doi:10.1103/PhysRevA.58.4824.

\bibitem[Sinatra \em{et~al.}(2002)Sinatra, Lobo, and Castin]{Sinatra2002a}
Sinatra, A.; Lobo, C.; Castin, Y.
\newblock The truncated Wigner method for Bose-condensed gases: Limits of
validity and applications.
\newblock {\em J. Phys. B} {\bf 2002}, doi:10.1088/0953-4075/35/17/301.

\bibitem[Isella and Ruostekoski(2006)]{Isella2006a}
Isella, L.; Ruostekoski, J.
\newblock Quantum dynamics in splitting a harmonically trapped Bose-Einstein
condensate by an optical lattice: Truncated Wigner approximation.
\newblock {\em Phys. Rev. A} {\bf 2006}, doi:10.1103/PhysRevA.74.063625.

\bibitem[Blakie \em{et~al.}(2008)Blakie, Bradley, Davis, Ballagh, and
Gardiner]{Blakie2008a}
Blakie, P.B.; Bradley, A.S.; Davis, M.J.; Ballagh, R.J.; Gardiner, C.W.
\newblock Dynamics and statistical mechanics of ultra-cold Bose gases using
c-field techniques.
\newblock {\em Adv. Phys.} {\bf 2008}, doi:10.1080/00018730802564254.

\bibitem[Martin and Ruostekoski(2010)]{Martin2010a}
Martin, A.D.; Ruostekoski, J.
\newblock Quantum and Thermal Effects of Dark Solitons in a One-Dimensional
Bose Gas.
\newblock {\em Phys. Rev. Lett.} {\bf 2010}, doi:10.1103/PhysRevLett.104.194102.

\bibitem[Polkovnikov(2010)]{Polkovnikov2010a}
Polkovnikov, A.
\newblock Phase space representation of quantum dynamics.
\newblock {\em Ann. Phys. (N.Y.)} {\bf 2010}, {\em 325},~1790--1852.

\bibitem[Opanchuk and Drummond(2013)]{Opanchuk2012a}
Opanchuk, B.; Drummond, P.D.
\newblock Functional Wigner representation of quantum dynamics of Bose-Einstein
condensate.
\newblock {\em J. Math. Phys.} {\bf 2013}, doi:10.1063/1.4801781.

\bibitem[Mathey \em{et~al.}({2014})Mathey, Clark, and Mathey]{Mathey}
Mathey, A.C.; Clark, C.W.; Mathey, L.
\newblock {Decay of a superfluid current of ultracold atoms in a toroidal
trap}.
\newblock {\em {Phys. Rev. A}} {\bf {2014}}, doi:10.1103/PhysRevA.90.023604.

\bibitem[Gross \em{et~al.}(2011)Gross, Est\`eve, Oberthaler, Martin, and
Ruostekoski]{gross_esteve_11}
Gross, C.; Est\`eve, J.; Oberthaler, M.K.; Martin, A.D.; Ruostekoski, J.
\newblock Local and spatially extended sub-Poisson atom-number fluctuations in
optical lattices.
\newblock {\em Phys. Rev. A} {\bf 2011}, doi:10.1103/PhysRevA.84.011609.

\bibitem[Plimak and Olsen({2014})]{Plimak}
Plimak, L.I.; Olsen, M.K.
\newblock {Quantum-field-theoretical approach to phase-space techniques:
Symmetric Wick theorem and multitime Wigner representation}.
\newblock {\em {Ann. Phys.}} {\bf {2014}}, {\em {351}},~{593--619}.

\bibitem[Martin and Ruostekoski({2012})]{Martin_bright}
Martin, A.D.; Ruostekoski, J.
\newblock {Quantum dynamics of atomic bright solitons under splitting and
recollision, and implications for interferometry}.
\newblock {\em {New J. Phys.}} {\bf {2012}}, doi:10.1088/1367-2630/14/4/043040.

\bibitem[Opanchuk \em{et~al.}({2012})Opanchuk, Egorov, Hoffmann, Sidorov, and
Drummond]{Opanchuk_inter}
Opanchuk, B.; Egorov, M.; Hoffmann, S.; Sidorov, A.I.; Drummond, P.D.
\newblock {Quantum noise in three-dimensional BEC interferometry}.
\newblock {\em {EPL}} {\bf {2012}}, doi:10.1209/0295-5075/97/50003.

\bibitem[Dujardin \em{et~al.}({2015})Dujardin, Uelles, and
Schlagheck]{Dujardin}
Dujardin, J.; Uelles, A.A.; Schlagheck, P.
\newblock {Elastic and inelastic transmission in guided atom lasers: A
truncated Wigner approach}.
\newblock {\em {Phy. Rev. A}} {\bf {2015}}, doi:10.1103/PhysRevA.91.033614.

\bibitem[Lewis-Swan \em{et~al.}(2015)Lewis-Swan, Olsen, and
Kheruntsyan]{Lewis-Swan}
Lewis-Swan, R.J.; Olsen, M.K.; Kheruntsyan, K.V.
\newblock On the interpretation of single stochastic trajectories of the Wigner
distribution.
\newblock {\em arXiv:1503.05647} {\bf 2015}.

\bibitem[Norrie \em{et~al.}(2006)Norrie, Ballagh, Gardiner, and
Bradley]{Norrie2006b}
Norrie, A.A.; Ballagh, R.J.; Gardiner, C.W.; Bradley, A.S.
\newblock Three-body recombination of ultracold Bose gases using the truncated
Wigner method.
\newblock {\em Phys. Rev. A} {\bf 2006}, doi:10.1103/PhysRevA.73.043618.

\bibitem[Szigeti \em{et~al.}(2009)Szigeti, Hush, Carvalho, and
Hope]{Szigeti2009a}
Szigeti, S.S.; Hush, M.R.; Carvalho, A.R.R.; Hope, J.J.
\newblock Continuous measurement feedback control of a Bose-Einstein condensate
using phase-contrast imaging.
\newblock {\em Phys. Rev. A} {\bf 2009}, doi:10.1103/PhysRevA.80.013614.

\bibitem[Hush \em{et~al.}(2013)Hush, Szigeti, Carvalho, and Hope]{Hush2013a}
Hush, M.R.; Szigeti, S.S.; Carvalho, A.R.R.; Hope, J.J.
\newblock Controlling spontaneous-emission noise in measurement-based feedback
cooling of a {Bose-Einstein} condensate.
\newblock {\em New J. Phys.} {\bf 2013}, doi:10.1088/1367-2630/15/11/113060.

\bibitem[Carmichael and Nha(2004)]{Carmichael_SED}
Carmichael, H.J.; Nha, H.
\newblock Vacuum fluctuations and the conditional homodyne detection of
squeezed light.
\newblock {\em J. Opt. B Quant. Semiclass. Opt.} {\bf
2004}, doi:10.1088/1464-4266/6/8/004.

\bibitem[Ruostekoski and Martin(2014)]{twabook}
Ruostekoski, J.; Martin, A.D.
\newblock {\em Quantum Gases: Finite Temperature and Non-Equilibrium Dynamics,
Cold Atoms Series};  Proukakis, N.P., Gardiner, S.A., Eds.;    Imperial College Press: London, UK,  2014.

\bibitem[Cattani \em{et~al.}(2013)Cattani, Gross, Oberthaler, and
Ruostekoski]{Cattani2013a}
Cattani, F.; Gross, C.; Oberthaler, M.K.; Ruostekoski, J.
\newblock Measuring and engineering entropy and spin squeezing in weakly linked
{Bose-Einstein} condensates.
\newblock {\em New J. Phys.} {\bf 2013}, doi:10.1088/1367-2630/15/6/063035.

\bibitem[Kippenberg and Vahala(2007)]{Kippenberg2007a}
Kippenberg, T.J.; Vahala, K.J.
\newblock Cavity Opto-Mechanics.
\newblock {\em Opt. Express} {\bf 2007}, {\em 15},~17172--17205.

\bibitem[Kippenberg and Vahala(2008)]{Kippenberg2008a}
Kippenberg, T.J.; Vahala, K.J.
\newblock Cavity Optomechanics: Back-Action at the Mesoscale.
\newblock {\em Science} {\bf 2008}, {\em 321},~1172--1176.

\bibitem[Meystre(2013)]{Meystre2013a}
Meystre, P.
\newblock A short walk through quantum optomechanics.
\newblock {\em Ann. Phys. (Leipzig)} {\bf 2013}, {\em 525},~215--233.

\bibitem[Aspelmeyer \em{et~al.}(2013)Aspelmeyer, Kippenberg, and
Marquardt]{Aspelmeyer2013a}
Aspelmeyer, M.; Kippenberg, T.J.; Marquardt, F.
\newblock Cavity Optomechanics.
\newblock {\em arXiv:1303.0733} {\bf 2013}.

\bibitem[{O'Connell} \em{et~al.}(2010){O'Connell}, Hofheinz, Ansmann, Bialczak,
Lenander, Lucero, Neeley, Sank, Wang, Weides, Wenner, Martinis, and
Cleland]{Oconnell2010a}
{O'Connell}, A.D.; Hofheinz, M.; Ansmann, M.; Bialczak, R.C.; Lenander, M.;
Lucero, E.; Neeley,~M.; Sank, D.; Wang, H.; Weides, M.; \emph{et al}.
\newblock Quantum ground state and single-phonon control of a mechanical
resonator.
\newblock {\em Nature} {\bf 2010}, {\em 464},~697--703.

\bibitem[Teufel \em{et~al.}(2011)Teufel, Donner, Li, Harlow, Allman, Cicak,
Sirois, Whittaker, Lehnert, and Simmonds]{Teufel2011a}
Teufel, J.D.; Donner, T.; Li, D.; Harlow, J.W.; Allman, M.S.; Cicak, K.;
Sirois, A.J.; Whittaker,~J.D.; Lehnert, K.W.; Simmonds, R.W.
\newblock Sideband cooling of micromechanical motion to the quantum ground
state.
\newblock {\em Nature} {\bf 2011}, {\em 475},~359--363.

\bibitem[Chan \em{et~al.}(2011)Chan, Alegre, Safavi-Naeini, Hill, Krause,
Gr\"{o}blacher, Aspelmeyer, and Painter]{Chan2011a}
Chan, J.; Alegre, T.P.M.; Safavi-Naeini, A.H.; Hill, J.T.; Krause, A.;
Gr\"{o}blacher, S.; Aspelmeyer,~M.; Painter, O.
\newblock Laser cooling of a nanomechanical oscillator into its quantum ground
state.
\newblock {\em Nature} {\bf 2011}, {\em 478},~89--92.

\bibitem[Brahms and Stamper-Kurn(2010)]{Brahms2010a}
Brahms, N.; Stamper-Kurn, D.M.
\newblock Spin optodynamics analog of cavity optomechanics.
\newblock {\em Phys.~Rev.~A} {\bf 2010}, doi:10.1103/PhysRevA.82.041804.

\bibitem[Gardiner(2009)]{GardinerStochastic}
Gardiner, C.
\newblock {\em Stochastic Methods}; Springer: Berlin, Germany,  2009.

\bibitem[Mekhov \em{et~al.}(2007)Mekhov, Maschler, and
Ritsch]{PhysRevLett.98.100402}
Mekhov, I.B.; Maschler, C.; Ritsch, H.
\newblock Cavity-Enhanced Light Scattering in Optical Lattices to Probe Atomic
Quantum Statistics.
\newblock {\em Phys. Rev. Lett.} {\bf 2007}, doi:10.1103/PhysRevLett.98.100402.

\bibitem[Ruostekoski \em{et~al.}(2009)Ruostekoski, Foot, and
Deb]{ruostekoski:170404}
Ruostekoski, J.; Foot, C.J.; Deb, A.B.
\newblock Light Scattering for Thermometry of Fermionic Atoms in an Optical
Lattice.
\newblock {\em Physical Review Letters} {\bf 2009}, doi:10.1103/PhysRevLett.103.170404.

\bibitem[Douglas and Burnett(2011)]{PhysRevA.84.053608}
Douglas, J.S.; Burnett, K.
\newblock Imaging of quantum Hall states in ultracold atomic gases.
\newblock {\em Phys. Rev. A} {\bf 2011}, doi:10.1103/PhysRevA.84.053608.

\bibitem[\L{}akomy \em{et~al.}(2009)\L{}akomy, Idziaszek, and
Trippenbach]{PhysRevA.80.043404}
\L{}akomy, K.; Idziaszek, Z.; Trippenbach, M.
\newblock Thermal effects in light scattering from ultracold bosons in an
optical lattice.
\newblock {\em Phys. Rev. A} {\bf 2009}, doi:10.1103/PhysRevA.80.043404.

\bibitem[Rist \em{et~al.}(2010)Rist, Menotti, and Morigi]{PhysRevA.81.013404}
Rist, S.; Menotti, C.; Morigi, G.
\newblock Light scattering by ultracold atoms in an optical lattice.
\newblock {\em Phys. Rev. A} {\bf 2010}, doi:10.1103/PhysRevA.81.013404.

\bibitem[Corcovilos \em{et~al.}(2010)Corcovilos, Baur, Hitchcock, Mueller, and
Hulet]{PhysRevA.81.013415}
Corcovilos, T.A.; Baur, S.K.; Hitchcock, J.M.; Mueller, E.J.; Hulet, R.G.
\newblock Detecting antiferromagnetism of atoms in an optical lattice via
optical Bragg scattering.
\newblock {\em Phys. Rev. A} {\bf 2010}, doi:10.1103/PhysRevA.81.013415.

\bibitem[Cordobes~Aguilar \em{et~al.}(2014)Cordobes~Aguilar, Ho, and
Ruostekoski]{AFMdetection}
Cordobes~Aguilar, F.; Ho, A.F.; Ruostekoski, J.
\newblock Optical Signatures of Antiferromagnetic Ordering of Fermionic Atoms
in an Optical Lattice.
\newblock {\em Phys. Rev. X} {\bf 2014}, doi:10.1103/PhysRevX.4.031036.

\bibitem[Goldwin \em{et~al.}(2014)Goldwin, Venkatesh, and O'Dell]{odell1}
Goldwin, J.; Venkatesh, B.P.; O'Dell, D.H.J.
\newblock Backaction-Driven Transport of Bloch Oscillating Atoms in Ring
Cavities.
\newblock {\em Phys. Rev. Lett.} {\bf 2014}, doi:10.1103/PhysRevLett.113.073003.

\bibitem[Samoylova \em{et~al.}(2015)Samoylova, Piovella, Robb, Bachelard, and
Courteille]{Samoylova:15}
Samoylova, M.; Piovella, N.; Robb, G.R.M.; Bachelard, R.; Courteille, P.W.
\newblock Synchronization of Bloch oscillations by a ring cavity.
\newblock {\em Opt. Express} {\bf 2015}, {\em 23},~14823--14835.

\bibitem[Horak and Ritsch(2001)]{Horak_singlephoton}
Horak, P.; Ritsch, H.
\newblock Manipulating a Bose-Einstein condensate with a single photon.
\newblock {\em Eur. Phys. J. D Atom. Mol. Opt.
Plasma Phys.} {\bf 2001}, {\em 13},~279--287.

\bibitem[Horak \em{et~al.}(2000)Horak, Barnett, and Ritsch]{Horak_beccavity}
Horak, P.; Barnett, S.M.; Ritsch, H.
\newblock Coherent dynamics of Bose-Einstein condensates in high-finesse
optical cavities.
\newblock {\em Phys. Rev. A} {\bf 2000}, doi:10.1103/PhysRevA.61.033609.

\bibitem[Javanainen and Yoo(1996)]{Javanainen1996a}
Javanainen, J.; Yoo, S.M.
\newblock Quantum Phase of a Bose-Einstein Condensate with an Arbitrary Number
of Atoms.
\newblock {\em Phys. Rev. Lett.} {\bf 1996}, {\em 76},~161--164.

\bibitem[Cirac \em{et~al.}(1996)Cirac, Gardiner, Naraschewski, and
Zoller]{Cirac1996a}
Cirac, J.I.; Gardiner, C.W.; Naraschewski, M.; Zoller, P.
\newblock Continuous observation of interference fringes from Bose condensates.
\newblock {\em Phys. Rev. A} {\bf 1996}, {\em 54},~R3714--R3717.

\bibitem[Castin and Dalibard(1997)]{Castin1997a}
Castin, Y.; Dalibard, J.
\newblock Relative phase of two Bose-Einstein condensates.
\newblock {\em Phys. Rev. A} {\bf 1997}, {\em 55},~4330--4337.

\bibitem[Ruostekoski and Walls(1998)]{RUO98}
Ruostekoski, J.; Walls, D.F.
\newblock Bose-Einstein condensate in a double-well potential as an open
quantum system.
\newblock {\em Phys. Rev. A} {\bf 1998}, {\em 58},~R50--R53.

\bibitem[Ruostekoski and Walls(1997)]{Ruostekoski1997a}
Ruostekoski, J.; Walls, D.F.
\newblock Nondestructive optical measurement of relative phase between two
Bose-Einstein condensates.
\newblock {\em Phys. Rev. A} {\bf 1997}, {\em 56},~2996--3006.

\bibitem[Javanainen(2010)]{javanainen10}
Javanainen, J.
\newblock Nonlinearity from quantum mechanics: Dynamically unstable
Bose-Einstein condensate in a double-well trap.
\newblock {\em Phys. Rev. A} {\bf 2010}, doi:10.1103/PhysRevA.81.051602.

\bibitem[Elliott \em{et~al.}(2015)Elliott, Kozlowski, Caballero-Benitez, and
Mekhov]{Elliott_QM}
Elliott, T.J.; Kozlowski, W.; Caballero-Benitez, S.F.; Mekhov, I.B.
\newblock Multipartite Entangled Spatial Modes of Ultracold Atoms Generated and
Controlled by Quantum Measurement.
\newblock {\em Phys. Rev. Lett.} {\bf 2015}, doi:10.1103/PhysRevLett.114.113604.

\bibitem[Corney and Milburn(1998)]{Corney1998a}
Corney, J.F.; Milburn, G.J.
\newblock Homodyne measurements on a Bose-Einstein condensate.
\newblock {\em Phys. Rev. A} {\bf 1998}, {\em 58},~2399--2406.

\bibitem[Dalvit \em{et~al.}(2002)Dalvit, Dziarmaga, and Onofrio]{Dalvit2002a}
Dalvit, D.A.R.; Dziarmaga, J.; Onofrio, R.
\newblock Continuous quantum measurement of a Bose-Einstein condensate: A
stochastic Gross-Pitaevskii equation.
\newblock {\em Phys. Rev. A} {\bf 2002}, doi:10.1103/PhysRevA.65.053604.

\bibitem[Lee \em{et~al.}(2012)Lee, Rist, and Ruostekoski]{Lee2012a}
Lee, M.D.; Rist, S.; Ruostekoski, J.
\newblock Bragg spectroscopic interferometer and quantum measurement-induced
correlations in atomic {Bose-Einstein} condensates.
\newblock {\em New J. Phys.} {\bf 2012}, doi:10.1088/1367-2630/14/7/073057.

\bibitem[Niedenzu \em{et~al.}(2013)Niedenzu, Sch\"{u}tz, Habibian, Morigi, and
Ritsch]{Niedenzu2013a}
Niedenzu, W.; Sch\"{u}tz, S.; Habibian, H.; Morigi, G.; Ritsch, H.
\newblock Seeding patterns for self-organization of photons and atoms.
\newblock {\em Phys. Rev. A} {\bf 2013}, doi:10.1103/PhysRevA.88.033830.

\bibitem[Mekhov and Ritsch(2009)]{Mekhov2009a}
Mekhov, I.B.; Ritsch, H.
\newblock Quantum Nondemolition Measurements and State Preparation in Quantum
Gases by Light Detection.
\newblock {\em Phys. Rev. Lett.} {\bf 2009}, doi:10.1103/PhysRevLett.102.020403.

\bibitem[Wiseman and Milburn(2010)]{Wiseman2010a}
Wiseman, H.; Milburn, G.
\newblock {\em Quantum Measurement and Control}; Cambridge University Press, Cambridge, UK,
2010.

\end{thebibliography}

\end{document}